\documentclass[runningheads]{llncs}

\usepackage{tabularx}
\usepackage{booktabs} 
\usepackage{tabulary}
\usepackage[utf8]{inputenc}
\usepackage[T1]{fontenc}
\usepackage{textcomp}
\usepackage{graphicx}
\usepackage{wrapfig} 
\usepackage{subcaption} 
\usepackage{float}%
\usepackage{xspace,tcolorbox}
\usepackage{tikz}
\usetikzlibrary{arrows.meta, positioning, shapes.geometric}
\usetikzlibrary{calc} 
\usepackage{caption}
\usepackage[normalem]{ulem}
\usepackage{hyperref} 
\usepackage[table]{xcolor}
\usepackage{xcolor}
\usepackage{colortbl}

\usepackage{makecell}

\usepackage{amsmath}
\usepackage{amssymb}
\usepackage{empheq}
\usepackage[mathscr]{euscript}
\DeclareSymbolFont{rsfs}{U}{rsfs}{m}{n}
\DeclareSymbolFontAlphabet{\mathscrsfs}{rsfs}
\usepackage{mathtools}
\usepackage{stmaryrd} 

\usepackage{dutchcal} 

\usepackage{pgfplots}
\pgfplotsset{compat=newest}

\usepackage[utf8]{inputenc}
\usepackage[T1]{fontenc}
\usepackage{textcomp,mathcomp}
\usepackage{xcolor,colortbl} 

\usepackage{boldline}

\usepackage{booktabs}

\usepackage{chngcntr} 

\newcommand{\halfcheckmark}{{$\checkmark$}\textsuperscript{\textcolor{black}{\kern-0.53em{\bf--}}}}

\usepackage{threeparttable}

\usepackage{booktabs}
\usepackage{nicematrix}
\usepackage{amsmath}

\usepackage{multirow}
\usepackage[ruled,linesnumbered]{algorithm2e}

\usepackage{breakcites}
\usepackage{enumitem}
\usepackage{stfloats}

\newcommand{\revisit}[1]{{\color{orange}#1}}

\newcommand{\eyasu}[1]{{\color{red!}{E: #1}}}
\newcommand{\howard}[1]{{\color{red!}{H: #1}}}

\usepackage{xargs} 
\usepackage[pdftex,dvipsnames]{xcolor}  
\usepackage[colorinlistoftodos,prependcaption,textsize=tiny]{todonotes}
\newcommandx{\commentt}[2][1=]{\todo[linecolor=red,backgroundcolor=red!25,bordercolor=red,#1]{#2}}

\usepackage[nocompress]{cite}
\usepackage[left=3.6cm, right = 3.6cm, top=3.6cm, bottom=3.6cm, asymmetric]{geometry}

\begin{document}

\title{BlowLive: Blow-Based Multi-Factor Biometrics with Liveness Detection and Revocability}

 \author{Eyasu Getahun Chekole, Howard Halim, Dani\"el Reijsbergen, Jianying Zhou}
\institute{Singapore University of Technology and Design}

%
\maketitle


\begin{abstract}

Biometric authentication systems are increasingly deployed in security-critical applications, yet existing physiological and behavioral biometrics suffer from fundamental limitations: 1) they are vulnerable to spoofing attacks due to unreliable liveness detection, 2) biometric templates may leak privacy-sensitive information 3) intra-user variability results in accuracy degradation, and 4) it is difficult to revoke physiological biometrics and safeguard them over long-term use. 
To address these challenges, we propose \emph{BlowLive}, a robust multi-factor biometric (MFB) 
framework that integrates blow-acoustic signals and facial biometrics as complementary behavioral and physiological modalities. BlowLive incorporates advanced spectral feature extraction and multimodal fusion techniques, achieving high authentication accuracy 
even for behavioral modalities. 
Instead of relying on conventional biometric approaches that utilize raw biometric templates for authentication, the proposed framework adopts a \emph{fuzzy-extractor–based biometric authentication scheme}, wherein stable cryptographic keys are derived from inherently noisy biometric inputs 
and subsequently used for authentication. 
To defend against playback, synthetic, and deepfake attacks, BlowLive further integrates a novel \emph{Doppler shift-based liveness detection} mechanism.  
We implement the complete BlowLive framework and experimentally evaluate 
its effectiveness using biometric data collected from 50 participants. The experimental results demonstrate high authentication accuracy ($99.56\%$ for blow-acoustics and $100\%$ for facial and fusion modalities), robust liveness detection ($99.46\%$ accuracy), strong template protection and revocability, non-invasiveness, and high usability. 

\end{abstract}


\section{Introduction}

Biometric authentication has become an indispensable component of modern identity verification systems, offering convenience and improved usability over traditional knowledge- and token-based authentication factors. Despite their widespread deployment across 
application areas, current biometric technologies still face fundamental limitations that hinder their reliability, robustness, and long-term security. These limitations span both 
physiological biometrics \cite{alsaadi2015physiological} (e.g., face, fingerprint, and iris) and behavioral biometrics \cite{finnegan2024utility} (e.g., voice, gait, keystroke, and touch dynamics), each presenting distinct challenges that remain insufficiently addressed in the literature and in practice.\
Physiological biometrics generally achieve higher authentication accuracy under controlled conditions, yet they are increasingly vulnerable to sophisticated spoofing and deepfake attacks \cite{wang2024deepfake, madan2023effect}. Advancements in generative AI now enable realistic synthetic faces, fingerprints, and iris textures, undermining the assumption that physiological traits are inherently difficult to replicate \cite{khan2025generative, makrushin2023survey}. 
Behavioral biometrics, in contrast, typically offer higher user convenience and resistance to 
a wide-range of spoofing attacks, but they generally suffer from lower accuracy, higher intra-user variability, and environmental sensitivity \cite{ayeswarya2024comprehensive}. 

A common and critical limitation across both biometric categories is the lack of reliable \textit{liveness detection}, i.e., detecting that the 
biometric sample comes from a physically present live person rather than a fake representation 
of prerecorded or deepfaked sample. 
Existing methods often rely on shallow heuristics, static image cues, or hand-crafted features that fail under adaptive attacks or when adversaries leverage deepfake technologies \cite{lu2018lippass, wu2019lvid}. 
Effective liveness detection must discriminate between genuine biometric samples and high-quality spoofs. However, current systems often fail to provide strong generalization across diverse attack types and biometric traits.  
Furthermore, many approaches \cite{zhang2022continuous, zhou2024securing, liu2022contactless} are ineffective for biometric modalities that lack the elaborate lip-motion patterns and vocal-tract dynamics resulting from speech, such as blow-acoustic biometrics. 
A further challenge lies in the security and long-term protection of biometric templates \cite{abdullahi2024biometric, lee2023reconstruction, srinivas2025enhanced}. Unlike passwords or cryptographic keys, biometric traits (especially physiological traits) cannot be changed once compromised. Consequently, biometric template leakage can be catastrophic, especially when biometric data is stored in plaintext or insufficiently protected formats. While template protection schemes and cancelable biometrics have been proposed, many suffer from degraded authentication accuracy, insufficient unlinkability, or impractical deployment requirements. Moreover, physiological biometrics are particularly non-revocable, making compromised templates effectively permanent.
Although there are prior works in biometric revocation or cancelable biometrics 
\cite{yang2022linear, hmani2024revocable, imran2025privacy, champaneria2025cancelable, belhocine2025medbioch, ragendhu2024cancelable, wang2024make}, these approaches primarily modify associated keys or user-specific parameters while the underlying biometric traits remain unchanged. As a result, they do not fully mitigate the security risks that arise when the original biometric 
trait is compromised.
Other major unmet requirements in most biometric systems are non-invasiveness 
and user convenience (usability) including seamless integration of different modalities 
and 
robustness in different postural modes, rapid execution, and minimal resource requirements, among others \cite{blowprint, ayeswarya2024comprehensive, kruzikova2022usable}. 
Together, these limitations highlight the necessity for next-generation biometric authentication frameworks that offer robust attack resistance, reliable liveness verification, secure and revocable template protection, and stable cryptographic integration. 

To overcome these limitations, we propose BlowLive, a novel 
MFB framework that integrate \textit{phone blowing-based acoustic signals} -- i.e., the audio signal produced by blowing onto the phone -- and \textit{facial data} as complementary behavioral and physiological biometric factors, respectively. In the former, we assume that 
the manner in which
users blow on their phone can produce distinctive acoustic patters that can serve as a unique behavioral biometric for user authentication. In the latter, we employ 
facial features captured 
\textit{simultaneously} when performing the phone blowing as a physiological authentication factor. BlowLive seamlessly integrates these two fundamentally different biometric factors, thereby satisfying modern multi-factor authentication requirements while significantly raising the security bar against impersonation and spoofing attacks. 


The proposed BlowLive framework incorporates a fuzzy-extractor (FE)-based authentication architecture that provides 
strong biometric template protection and supports biometric revocation in the event of compromise. Rather than relying on raw biometric templates, we derive stable, cryptographically strong keys using a fuzzy extractor, ensuring that authentication is performed using privacy-preserving representations instead of sensitive biometric features. We also employ different feature extraction and embedding techniques in this scheme. For the blow-acoustic modality, we employ Gammatone Frequency Cepstral Coefficients (GFCC) \cite{gfcc}, a robust feature extraction technique capable of capturing fine-grained gammatone and broadband characteristics of user-specific blowing patterns. 
The GFCC features extracted from each recording are passed through a Convolutional Neural Network (CNN) model to produce discriminative blow-acoustic embeddings. 
These embeddings can then be combined with the facial embeddings produced by the FaceNet model through
score-level or feature-level fusion. 
New cryptographic keys are generated from the fused embeddings if the hamming distance ($HD$) between the query embeddings and the enrolled ones falls within an acceptable threshold. Furthermore, this scheme enables revocation of both the biometric templates and the derived cryptographic keys in case of compromise, thus mitigating a longstanding limitation of biometric systems. 
By simply enrolling new blow-acoustic patterns, users can obtain fresh biometric identities, thereby 
mitigating one of the longstanding limitations of traditional biometric systems. 

To further enhance the system against a wide-range of spoofing attacks, 
we incorporate a Doppler shift-based \textit{liveness detection} module, specifically designed to mitigate playback, synthetic, and deepfake-based attacks targeting acoustic channels. 
Unlike speech-based Doppler analysis, which can rely on structured harmonic content, blow-acoustic signals lack voiced components, making Doppler-based analysis significantly more challenging. To overcome this, we develop a specialized Doppler processing pipeline using advanced time-frequency feature extraction and hybrid descriptors to capture the air turbulence and micro-motion dynamics generated during genuine phone blowing interactions. 
This 
module offers robust protection against high-fidelity audio playback, AI-generated blowing patterns, and other sophisticated attack vectors. 

We evaluate the effectiveness of the proposed framework by collecting blow-acoustic and facial data from $50$ participants. 
The proposed FE-based authentication scheme achieves accuracies of $99.56\%$, $100\%$, $99.95\%$ and $100\%$ for the blow-acoustic, facial, score-level fusion, and feature-level fusion modalities, respectively. Notably, we achieve nearly $100\%$ accuracy even in the behavioral modality, which is traditionally characterized by instability and lower discriminative power. 
The proposed liveness detection module also achieves $99.42\%$ accuracy, demonstrating strong effectiveness in distinguishing live blow-acoustics from playback signals. The proposed framework also inherently achieves the other security requirements mentioned previously (i.e., non-invasiveness, usability, MF support, and minimal resource requirements).

A key contribution of BlowLive is its ability to achieve \emph{high accuracy} even in the behavioral modality, which is traditionally characterized by instability and lower discriminative power. This improvement is achieved by our advanced spectral analysis and feature extraction pipeline on the blow-acoustic signals.

Collectively, these contributions establish BlowLive as a comprehensive, secure, and revocable multi-factor biometric authentication solution. By addressing accuracy limitations, spoofing threats, liveness detection gaps, biometric template security, and biometric revocability (both biometric template and the associated cryptographic key), all within a unified framework, the proposed framework advances the state-of-the-art in biometric authentication and provides a practical path forward for deploying high-assurance biometric security in adversarial environments.

In general, this work makes the following key contributions:
\begin{itemize}
    \item 
    We design and implement BlowLive, a 
    robust and highly 
    effective MFB framework that seamlessly integrates blow-acoustic signals and facial data as complementary behavioral and physiological modalities. 
    \item 
    We achieve nearly 100\% authentication accuracy 
    by employing an advanced feature extraction technique, namely GFCC, on blow-acoustic data.
    \item The introduced FE-based biometric authentication strengthens template security and enhances user privacy by removing sensitive biometric templates while also providing biometric revocability both at template and key levels. 
     \item 
    We propose a novel Doppler-shift-based liveness detection module 
    that accurately detects playback, synthetic, and deepfake blow-acoustic signals. 
     \item 
    We rigorously evaluate the accuracy and performance of BlowLive 
    using high-quality blow-acoustic and facial 
    data collected from 50 participants.
\end{itemize}

\section{Related Works} \label{sec:related_work}

In this section, we review prior works that are most closely aligned with our work. A comparative summary of these works across multiple evaluation criteria is also provided in Table \ref{table_comp_related_works}.

\textbf{Multi-Factor Biometric Schemes.}
Chauhan et al. \cite{chauhan2017breathprint} evaluated sniffing, normal, and deep breathing patterns for user identification, reporting a true positive rate (TPR) 
of up to 94\%, but the reliance on only three behaviors limits scalability and adaptability, with modest overall accuracy. 
Al-Wasy et al. \cite{al2017multimodal} proposed a multimodal biometric system using a deep learning approach that integrates facial features and both left and right irises, employing a fusion module at the score and rank levels. 
Aizi et al. \cite{aizi2022score} and Srivastava et al. \cite{srivastava2022match} proposed multimodal biometric systems combining fingerprint-iris and finger-knuckle-iris modalities, respectively, using score- and match-level fusion. However, these 
systems exclusively rely on physiological biometrics, without incorporating behavioral modalities. 
El Rahman et al. \cite{rahman2020multimodal} proposed a hybrid approach combining ECG and fingerprint biometrics using multiple fusion strategies. While the combination of physiological and behavioral modalities aims to enhance security and robustness, the system remains somewhat invasive, requiring users to directly scan the fingerprint while additional equipment outside smartphone is required to capture the ECG signals.
Mahfouz et al. \cite{mahfouz2024m2auth} proposed a multimodal behavioral authentication system using feature-level fusion of touch gestures, keystroke dynamics, and accelerometer data, but reliance on behavioral traits alone can introduce variability due to user state and environmental factors, affecting accuracy and robustness. Lee et al. \cite{lee2021advanced} developed an authentication method for IoT devices that combined touch and motion data from smartphones and smartwatches. However, the requirement of an external wearable limits convenience and applicability when such devices are unavailable. 
Wu et al. \cite{wu2022echohand} proposed an authentication system that utilizes hand geometry features and acoustic sensing technique, whereas Zhou et al. \cite{zhou2018echoprint} proposed an authentication system that utilizes facial landmarks and acoustic sensing. Both systems utilize echoes as an acoustic features. 
Although both systems make effective use of echo-based acoustic features, they do not provide built-in liveness detection or revocation capabilities. 
Koffi et al. \cite{koffi2023voice} highlighted that, among the various biometrics modality combinations, voice and face biometrics stand out for their balance of robustness, security, and usability. Particularly due to their inherent support for liveness detection and minimal of intrusiveness. Building upon these insights, our proposed implementation addresses the aforementioned challenges by introducing a novel fusion of physiological and behavioral modalities that ensures improved security, usability, while maintaining high accuracy. 
Halim et al.\ \cite{blowprint} proposed a multimodal biometric authentication system using blow-based acoustic signals and facial features. However, their approach does not 
employ advanced feature extraction techniques, resulting in lower overall accuracy compared to ours as demonstrated in Section 
\ref{sec:experimental_evaluation}, and lacks
liveness detection and 
 revocability. 




\textbf{Liveness Detection.} Liveness detection has become a critical component of biometric systems, especially with the rise in spoofing attacks using high-quality replay or synthetic inputs. Various techniques have been proposed for different 
contexts. In this work, we focus on Doppler shift-based liveness detection, particularly as applied to acoustic and motion-driven biometric systems. Lu et al. \cite{lu2018lippass} proposed Lippass, a lip-reading-based user authentication systems that uses Doppler shifts of acoustic signals to verify liveness through lip movement, achieving 90.21\% user identification accuracy and 93.1\% liveness detection accuracy. Wu et al. \cite{wu2019lvid} proposed LVID, a multimodal biometric authentication system that integrates lip movements with voice signals to enable liveness detection. Their method achieves 95\% authentication accuracy and 93.47\% replay-attack detection accuracy. Zhang et al. \cite{zhang2022continuous} leverages smartphone speakers and microphones as Doppler radars to detect speech liveness by tracking articulatory motions, achieving over 99\% detection accuracy and around 1\% EER (equal error rate). 
Zhou et al. \cite{zhou2024securing} proposed FaceLip, a mobile face-liveness detection system that captures unforgeable lip-motion patterns through Doppler shifts, achieving 96.6\% accuracy and an EER of 6.99\%. While these approaches demonstrate strong effectiveness in liveness detection, 
they cannot be directly applied to blow-acoustic biometrics, as blowing lacks both lip-motion patterns and vocal-tract dynamics.
Liu et al. \cite{liu2022contactless} developed a contactless breathing-airflow sensing system that detects Doppler shifts using a smartphone's microphone and speaker, reporting 74 -- 98\% detection accuracy across diverse sensing conditions. 
However, the system is 
not suitable for short and dynamic blowing behaviors required for biometric authentication.



\textbf{Revocation.}
Revocation is a critical feature in biometrics that enhances the security of biometric systems by revoking biometric templates and/or cryptographic keys derived from biometric inputs. In this regard, we focus on revocation of biometrics using fuzzy extractors. Dodis et al. \cite{dodis2004fuzzy} introduce fuzzy extractors, constructs that derive strong cryptographic keys from noisy biometric data while ensuring reconstruction under slight variations. Boyen \cite{boyen2004reusable} develops reusable fuzzy extractors that remain secure across multiple enrollments. Canetti et al. \cite{canetti2016reusable} further optimize these designs for low-entropy biometric sources. These primitives form the basis for revocability in biometric systems, enabling template and cryptographic key renewal.
Yang et al. \cite{yang2022linear} proposed a cancelable biometric template protection for fingerprints that uses linear convolutional with user-dependent helper vectors, to create a unlinkable, non-invertible, and revocable template. However, the linearity of the transformation makes the scheme potentially vulnerable to inversion or correlation attacks if the helper vectors are compromised. 
Hmani et al. \cite{hmani2024revocable} proposed a revocable crypto-key regeneration scheme based on face biometrics, where a shuffling permutation is applied using a user-specific shuffle key $S$ during enrollment. Revocation requires issuing a new shuffle key $S'$, making the scheme dependent on user-provided keys for template protection. 
Imran et al. \cite{imran2025privacy} and proposed a hybrid cancellable fingerprint template generation method, combining minutiae translation, Möbius transformation, and permutation to achieve revocability. Champaneria et al. \cite{champaneria2025cancelable} introduced a fingerprint template protection technique using raster scan, affine transformation, and user-specific key sets to construct secure and diverse templates. Although both approaches achieve revocability and non-invertibility, their security relies on the user providing or managing the secret key, which may limit usability and introduces dependency on correct key handling. Belhocine et al. \cite{belhocine2025medbioch} proposed MedBioCh, a revocable biometric template scheme that uses a multi-layered chaotic approach, by projecting biometric data into the projected space created by the chaotic system. Revocation can be achieved by modifying the parameters of the chaotic system.
Ragendhu et al. \cite{ragendhu2024cancelable} proposed a cancelable biometric template using user-specific random vectors and median filtering to generate multiple transformed feature representations, storing only their pairwise distances to preserve irreversibility and support revocation. 
Wang et al. \cite{wang2024make} proposed CanFG, a virtual identity generation framework that produces protected face images embedding a transformed identity while enabling cancelability and revocability. In both approaches, revocation is achieved by generating a new set of random vectors or virtual identities, allowing the template or protected image to be replaced without exposing the original biometric data. However, these methods are vulnerable to inference attacks and residual visual similarity, respectively.



 While the aforementioned state-of-the-art approaches \cite{yang2022linear, hmani2024revocable, imran2025privacy, champaneria2025cancelable, belhocine2025medbioch, ragendhu2024cancelable, wang2024make} 
achieve revocability 
mainly by modifying keys or user-specific parameters, the underlying 
biometric data remains fixed. 




\section{Threat Model and Requirements}
\subsection{Threat Model}
In our threat model, we consider an adversary attempting to compromise the BlowLive biometric framework through spoofing or playback-based attacks. 
In particular, we assume the following capabilities of the adversary. 
\begin{itemize}
  \item 
  The phone microphone is not compromised 
  and it produces the expected biometric data from blow-acoustic input.
  \item \textbf{Playback attack:} The adversary may record a genuine user's blowing 
  gesture and replays it via a speaker to bypass authentication. 
  \item \textbf{Stolen sample attack:} The adversary may obtain an authentic blow sample (stored or leaked) of a user and 
  reuses it to fool the system  during authentication. 
  \item \textbf{Deepfake attack}: The adversary may use AI-generated blow-acoustic data 
  to bypass the liveness detection. 
\end{itemize}


\subsection{Security Requirements}
\label{sec:requirements}
In the following, we outline the desirable security requirements for biometric authentication and we use 
them as evaluation criteria for evaluating existing and proposed biometric techniques.  

\textbf{(R1)} \emph{Accuracy}: A behavioral biometric technique should demonstrate high accuracy to effectively minimize the risk of impersonation attacks (i.e., by having low false positives), while also enhancing user convenience by lowering the number of required authentication attempts (i.e., by having low false negatives). 

\textbf{(R2)} \emph{Liveness Detection}: With the rise of AI-based deepfaking, 
biometric techniques should support detection of liveness, i.e., that the data originates from a living person instead of a spoofed representation, to verify authenticity. 

\textbf{(R3)} \emph{Revocability}: In biometrics, revoking biometric templates and/or the 
derived cryptographic keys 
is 
critical to enhance security in case of compromise. 

\textbf{(R4)} \emph{Resilience Against Known Attacks}: A biometric technique should demonstrate high resilience against certain known attacks, such as 
database forging, replay, adversarial ML, biometric replication, and template inversion 
attacks. 

\textbf{(R5)} \emph{Non-Invasiveness}:
 To address hygiene- and certain privacy-related concerns, behavioral biometric techniques should be performed in a touchless manner and only in a few seconds. 

\textbf{(R6)} \emph{Usability}:
One of the primary challenges in behavioral biometrics and MFB is usability. In this regard, the key usability factors include:
\begin{itemize}
\item \emph{User convenience}: The system should be intuitive (simple to use) and non-intrusive (not causing discomfort). 
\item \emph{Response time}: Authentication should be fast and seamless.
\item \emph{Robustness across posture modes}: the system's accuracy should not require a specific posture 
(e.g., standing or sitting).
\item \emph{Seamless integration}: 
Combining multiple biometric techniques in a single authentication attempt (one-shot authentication) improves user experience by minimizing the need for separate actions and reducing processing time. 
\end{itemize}

 \textbf{(R7)} \emph{MFB Support}:
In light of the inherent limitations of single-factor biometrics, 
biometric authentication techniques should incorporate multiple biometric modalities, and particularly both
behavioral and physiological factors.
 
\textbf{(R8)} \emph{Low Resource Requirements}: 
Biometric authentication should be conducted using minimal resources, i.e., the computational resources and sensors available on a regular smartphone, without requiring additional hardware. 

\section{BlowLive: Proposed Technique}

\begin{wrapfigure}{r}{0.42\textwidth}
    \centering  
    \includegraphics[width=0.42\textwidth]{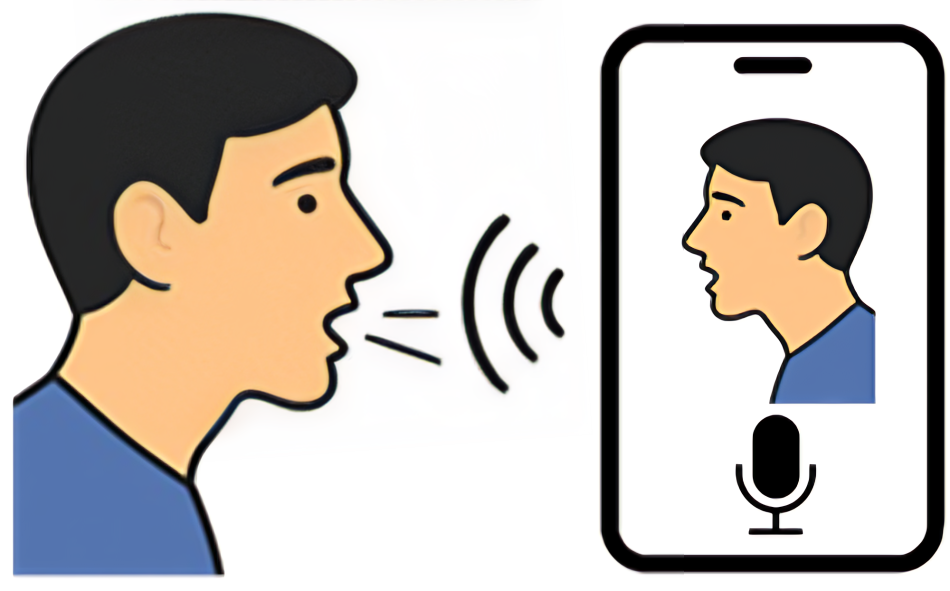}  
    \caption{Illustration of BlowLive}
    \label{fig:blowlive_illustration}
\end{wrapfigure}

This section provides a detailed description of \emph{BlowLive}, a novel 
MFB framework that effectively authenticates users based on their phone-blowing behaviors and facial features while 
also achieving liveness detection and biometric revocability. Specifically, the proposed technique employs phone blowing acoustic patterns and facial data as behavioral and physiological biometric authentication factors, respectively, which are simultaneously captured while blowing on the phone, as illustrated in Figure \ref{fig:blowlive_illustration}. 
 In FE-based authentication, a cryptographic key is first derived from the biometric data using a fuzzy extractor, and this key is subsequently used for user authentication. This scheme also enables secure revocation of the derived cryptographic key and, in the case of behavioral biometrics, the associated biometric template. To defend against biometric spoofing attacks, the proposed framework further incorporates a Doppler-shift based liveness detection 
technique, ensuring that the biometric input originates from a live and genuine user. 

\subsection{Workflow}
\begin{figure*}[!htbp]
    \centering  
     \includegraphics[width=\linewidth]{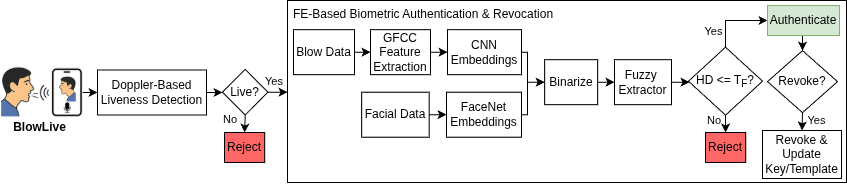} 
    \caption{A high-level workflow of BlowLive}
    \label{fig_blowlive_architecture}
\end{figure*}

The end-to-end workflow of BlowLive is illustrated in Figure \ref{fig_blowlive_architecture}. 
The workflow begins when the user interacts with the BlowLive application 
and simultaneously records the user's blow-acoustic signal and facial features. 
These raw inputs are forwarded to the liveness detection 
module, where the system verifies the liveness of the input. 
If the liveness check fails, the process terminates with a rejection. 
Otherwise, it proceeds 
to the FE-based authentication stage. 
Details of each process are provided in the following sections.

\subsection{Data Acquisition} \label{sec:data_acquisition}
To validate our proposed technique, we developed a proof-of-concept Android application, called \emph{BlowLive}, which captures a user's phone-blowing acoustic signals and facial features concurrently. 
The application also incorporates an imperceptible $20~\mathrm{kHz}$ ultrasonic emission. The interaction of this signal with the user’s blow produces a Doppler shift that is recorded by the device microphone, enabling the capture of fine-grained motion features for subsequent liveness analysis.




\subsection{Feature Extraction} \label{sec:feature_extraction}

In this section, we 
provide an overview of the feature extraction techniques 
used in the liveness detection (cf. Section \ref{sec:liveness_detection}) 
and FE-based authentication (cf. Section \ref{sec:feba_scheme}) 
modules. 

\subsubsection{\textbf{GFCC-Based Feature Extraction}} \label{sec:GFCC_feature_extraction}


To facilitate the extraction of meaningful features from the blow-acoustic signals, we implement an audio processing pipeline that computes GFCC representations. The audio waveform is segmented into frames of \emph{50 ms} with a \emph{20 ms} hop and transformed into 
GFCC by applying a 24-channel gammatone filterbank and computing the corresponding cepstral representation. After logarithmic compression, a discrete cosine transform (DCT) is applied to the 24 filter outputs, and the first 13 coefficients are retained:
\begin{equation}\label{eq:gfcc_coefficient_vector}
\mathrm{g}(t_n) = \mathrm{DCT}\big(\log\mathrm{G}(t_n)\big),
\end{equation}
where $\mathrm{G}(t_n)$ denotes the 24-dimensional gammatone filterbank energy vector at frame $t_n$, $\log \mathrm{G}(t_n)$ represents its log-compressed energies, and $\mathrm{g}(t_n)$ corresponds to the resulting 13-dimensional static GFCC coefficient vector.

To capture short-term temporal dynamics, first-order delta coefficients $\Delta \mathrm{g}(t_n)$ are computed using a 2-frame symmetric regression window. The final per-frame representation is obtained by concatenating the static and delta coefficients:
\begin{equation}
\mathrm{f}(t_n) = \big[\mathrm{g}(t_n),\; \Delta \mathrm{g}(t_n)\big],
\end{equation}
where $\mathrm{f}(t_n)$ denotes the resulting 26-dimensional GFCC feature vector for frame $t_n$, representing both the spectral characteristics and short-term temporal evolution of the blow-acoustic signal.


\subsubsection{\textbf{GFCC-CNN 
Feature Extraction}} \label{sec:gfcc_cnn}

To remove temporal dependencies in the GFCC sequences and produce stable embeddings that no longer behave as time series data, we develop a 
CNN-based model. 
This allows the protocol to learn discriminative embeddings from the GFCC representations of blow-acoustic signals. The model takes as input the GFCC sequence of a blow-acoustic recording, consisting of 26 channels (13 static + 13 delta) across all frames. The sequence is processed by a stack of 1D convolutional layers with batch normalization and Rectified Linear Unit (ReLU) \cite{relu} 
activation, followed by temporal mean-pooling: 

\[
\mathrm{h} = \text{MeanPool}(\text{ConvStack}(\mathrm{X})) \in \mathbb{R}^{F},
\]
where $\mathrm{h} \in \mathbb{R}^{F}$ denotes the representation of the pooled characteristics and $F$ denotes the number of channels output by the convolutional stack. 

This temporal pooling aggregates information across all frames and removes frame-wise temporal dependencies, producing a fixed-length representation independent of the number of frames. The pooled vector is then projected to a 128-dimensional embedding and L2-normalized:
\[
\mathrm{f}_\text{blow}(\mathrm{X}) = \frac{\text{FC}_2(\mathrm{ReLU}(\text{FC}_1(\mathrm{h})))}{\|\text{FC}_2(\mathrm{ReLU}(\text{FC}_1(\mathrm{h})))\|_2} \in \mathbb{R}^{128},
\]

where $\mathrm{f}_{\text{blow}}(\mathrm{X}) \in \mathbb{R}^{128}$ represents the final normalized blow-acoustic embedding, and $\mathrm{ReLU}(\cdot)$ is the 
ReLU activation function. 

The fully connected (FC) layers consist of $\text{FC}_1: \mathbb{R}^{F} \rightarrow \mathbb{R}^{256}$, which expands the pooled representation, followed by $\text{FC}_2: \mathbb{R}^{256} \rightarrow \mathbb{R}^{128}$, which projects it into the embedding space prior to normalization. 
The network is trained for $50$ epochs using the Adam optimizer (learning rate $1 \times 10^{-4}$, batch size $32$) with a triplet loss, producing discriminative 128-dimensional blow-acoustic feature vectors for both enrollment and authentication.

\subsubsection{FaceNet-Based Facial Feature Extraction}\label{sec:facenet_extraction}

To extract facial features, we adopt the Inception-ResNet v1 backbone from FaceNet using the \texttt{facen\-e\-t-\-p\-y\-torch} implementation \cite{esler_facenetpytorch} with pre-trained VGGFace2 weights. 
Given an input face image $I$, the backbone generates a 512-dimensional embedding:
\[
\mathrm{h} = \text{Backbone}(I) \in \mathbb{R}^{512},
\]
where $\mathrm{h} \in \mathbb{R}^{512}$ denotes the high-level facial embedding generated by the pretrained backbone. 
This embedding is then projected to a 128-dimensional feature vector using a small fully connected projection head with a ReLU activation:
\[
\mathrm{f}(I) = \mathrm{ReLU}(\text{Proj}(\mathrm{h})) \in \mathbb{R}^{128},
\]
where $\text{Proj}(\cdot)$ denotes the projection layer mapping $\mathrm{h}$ to a lower-dimensional space, and $\mathrm{f}(I) \in \mathbb{R}^{128}$ represents the resulting 128-dimensional facial feature vector. 

Although the backbone is pre-trained, the network (backbone + projection head) is further trained on our dataset using a triplet loss function for $50$ epochs with the Adam optimizer (learning rate $1 \times 10^{-4}$, batch size $32$). The resulting vector $\mathrm{f}(I)$ is used as the facial representation feature vector.

\subsection{Liveness Detection}\label{sec:liveness_detection}

To ensure that the proposed BlowLive authentication system is resistant to playback and synthetic audio attacks, we design a Doppler shift-based liveness detection module that verifies liveness of the user's biometric input. 
Unlike voice-based Doppler biometrics that exploit 
frequency variations arising from vocal cord vibrations, blow-acoustic biometrics rely on airflow turbulence generated during phone blowing. Accordingly, we exploit fine-grained frequency shifts from emitted ultrasonic signals introduced by the phone blowing airflow patterns relative to the microphone to verify the presence of a live human input. These Doppler signatures are difficult to replicate with pre-recorded or synthetic signals.
To verify liveness, we analyze the fine-grained Doppler frequency shifts imposed on the emitted ultrasonic signals by the 
phone blowing airflow patterns as they move relative to the microphone. 
The proposed liveness detection module consists of four main stages: (i) signal preprocessing, (ii) feature extraction, (iii) hybrid score computation, and (iv) 
threshold computation. 

\subsubsection{\textbf{Signal Preprocessing}}\label{sec:signal_preprocessing}

As noted above, the phone continuously emits an ultrasonic tone (e.g., $20~\mathrm{kHz}$) while the user blows toward the microphone. This ultrasonic tone serves as a reference carrier for subsequent frequency-shift analysis. As a result, the recorded audio contains both the emitted ultrasonic component and its reflections modulated by the user’s blowing motion.
As such, we preprocess the data to isolate the Doppler-bearing component and transform it into a form suitable for downstream analysis, and subsequently integrate it into the hybrid-score-based liveness detection framework. This preprocessing pipeline consists of the following main steps:

     \paragraph{Band-Pass Filter.} The raw audio $x(t)$ is first filtered using a fourth-order band-pass filter $H(f)$ centered around the emitted ultrasonic band (e.g., $19.5-20.5~\mathrm{kHz}$). This step removes all audible content and environmental noise while preserving only the frequency region where Doppler shifts occur. The resulting band-pass filtered signal $x_{\mathrm{bp}}(t)$, extracted around the emitted 
     tone, is computed as:
    \begin{equation}\label{eq:bandpass_filter}
    x_{\mathrm{bp}}(t) = H(f) \cdot x(t).
    \end{equation}
    \paragraph{Short-Time Fourier Transform (STFT) \cite{gao2021octonion}.} The filtered signal is transformed into a time-frequency representation to track how the energy peak around the ultrasonic carrier shifts over time. This enables frame-wise estimation of the instantaneous peak frequency $f_{\mathrm{peak}}(t_n)$. Let $Z(f_k,t_n)$ denote the complex STFT value at frequency $f_k$ and time frame $t_n$. For each frame, we identify 
    the $f_{\mathrm{peak}}(t_n)$ near the ultrasonic band (e.g., $19.5-20.5~\mathrm{kHz}$) at time frame $t_n$, i.e.,
    \begin{equation}
    f_{\mathrm{peak}}(t_n) = \underset{f_k \in [19.5,20.5]\text{kHz}}{\text{argmax}}
    |Z(f_k,t_n)|,
    \end{equation}
    \paragraph{Doppler Shift Extraction.} The instantaneous Doppler frequency shift $\Delta f$ (i.e., the deviation of the peak frequency from the original emitted tone $f_0$ caused by airflow motion during a blow) at time frame $t_n$ is computed as:
    \begin{equation}\label{eq:doppler_frequency_shift}
    \Delta f(t_n) = f_{\mathrm{peak}}(t_n) - f_0,
    \end{equation}
    where $f_0 = 20~\mathrm{kHz}$ is the emitted ultrasonic tone. 
    \paragraph{Doppler Envelope Computation.} 
    Finally, the Doppler envelope at time frame $t_n$ (i.e., $e(t_n)$), which captures the energy variation induced by airflow turbulence, is extracted as follows:
    \begin{equation}\label{eq:etn}
    e(t_n) = \sqrt{\frac{1}{K}\sum_{k=1}^{K} |Z(f_k,t_n)|^2},
    \end{equation}
    where $K$ is the number of frequency bins in the Doppler band (e.g., 19.5-20.5 kHz) used in the STFT. 

\subsubsection{\textbf{GFCC Feature Extraction}}\label{sec:liveness_gfcc_feature_extraction}
To characterize the broadband spectral structure of the blowing activity, we further extract and compute 
features using GFCC (see Section \ref{sec:GFCC_feature_extraction} for details). The band-pass filtered signal $x_{\mathrm{bp}}(t)$, obtained from the signal preprocessing stage (cf. Eq. \eqref{eq:bandpass_filter}), is passed through a $32$-channel gammatone filterbank, followed by logarithmic compression. 
The first $13$ cepstral coefficients are then computed for each frame using Eq. \eqref{eq:gfcc_coefficient_vector}, i.e., $\mathrm{g}(t_n) = \mathrm{DCT}\big(\log\mathrm{G}(t_n)\big)$. Using the resulting GFCC coefficient vector $\mathrm{g}(t_n)$ (derived from Eq. \eqref{eq:gfcc_coefficient_vector}), we compute the frame-averaged GFCC energy ($E_{\mathrm{GFCC}}$) across $N$ cepstral coefficients 
in the following way:
\begin{equation}\label{eq:egfcc}
E_{\mathrm{GFCC}} = \frac{1}{N}\sum_{n=1}^{N} \|\mathrm{g}(t_n)\|_2.
\end{equation}

\subsubsection{\textbf{Hybrid Liveness Scoring}}\label{sec:hybrid_liveness_scoring}

In the proposed liveness detection module, we employ a hybrid decision score that integrates three complementary acoustic-Doppler features extracted from the user’s blowing signal. These features capture distinct physical characteristics of legitimate human blowing acoustics and provide strong discriminative power against playback and synthetic attacks. 
The reason is that
playback and synthetic attacks typically produce abnormal Doppler profiles and attenuated airflow spectral components.

    \paragraph{GFCC Energy} ($E_{\mathrm{GFCC}}$). Represents the average spectral energy of the 
    GFCC, which model the perceptually relevant sub-band structure of the ultrasonic blowing signal. Genuine blowing typically produces a richer and more dynamically modulated cepstral pattern than replayed or artificially generated signals. Therefore, $E_{\mathrm{GFCC}}$ is computed using Eq. \eqref{eq:egfcc}, i.e., $E_{\mathrm{GFCC}} = \frac{1}{N}\sum_{n=1}^{N} \|\mathrm{g}(t_n)\|_2$, and used as an essential factor in the hybrid liveness score computation. 
    \paragraph{Doppler RMS Envelope} ($E_{\mathrm{RMS}}$). 
    Captures the temporal magnitude of motion-induced frequency deviations around the $20~kHz$ carrier tone. Because human blowing generates non-stationary airflow with varying velocity components, its Doppler envelope exhibits characteristic fluctuations that are difficult to reproduce with loudspeakers, and it is computed as:
    \begin{equation}\label{eq:erms}
        E_{\mathrm{RMS}} = \sqrt{\frac{1}{N}\sum_{n=1}^{N} e(t_n)^2},
    \end{equation}
     where $N$ is the number of time frames in the Doppler envelope sequence and $ e(t_n)$ is computed using Eq. \eqref{eq:etn}.
     \paragraph{Doppler Max} ($D_{max}$). Measures the largest absolute Doppler frequency shift $\Delta f(t_n)$ across all time frames, reflecting the highest transient velocity produced by the user’s blowing. Authentic blows produce natural peak variations associated with turbulent airflow, whereas attacks generally yield lower and more uniform Doppler shifts. For $n$ time frame indices, $D_{max}$ is computed as follows:
    \begin{equation}\label{eq:dmax}
        D_{max} = \max_n |\Delta f(t_n)|,
    \end{equation}
     where $\Delta f(t_n) = f_{\mathrm{peak}}(t_n) - f_0$ (derived from. Eq. \eqref{eq:doppler_frequency_shift}) and $f_0 = 20~\mathrm{kHz}$ is the emitted ultrasonic tone.  

\paragraph{Hybrid Scores.} The three features described above 
(i.e., $E_{GFCC}$, $RMS_e$ and $D_{max}$) are combined through a weighted linear fusion to form a robust hybrid liveness score that enhances the separability between legitimate and attack samples while remaining computationally lightweight. These features together provide spectral, motion, and velocity cues, forming a highly discriminative representation of live blowing versus spoofing (replay and synthetic) attacks. Therefore, the hybrid score $S_{hyb}$ is computed as follows: 
\begin{equation}
S_{hyb} = \alpha\, E_{\mathrm{GFCC}} + \beta\, E_{\mathrm{RMS}} + \gamma\, D_{max},
\end{equation}
where $(\alpha,\beta,\gamma)$ are the weighting coefficients used to control the relative importance of the three features and are chosen empirically. In our setting, they are assigned the values $(0.5,0.3,0.2)$, respectively. 

\subsubsection{\textbf{Threshold Computation}}\label{sec:threshold_computation}

We evaluate the liveness detection module using both global and per-user (dynamic) decision thresholds. In the global setting, a single threshold is computed from all legitimate training samples collected across all participants. The intuition behind this approach is to model population-level genuine behavior and apply a uniform acceptance criterion at runtime. In the per-user setting, an individualized threshold is computed for each enrolled user, based on the assumption that each user’s legitimate blowing behavior follows a distinct statistical distribution and may therefore benefit from a personalized acceptance boundary. The effectiveness of the two settings is validated empirically (cf. Section \ref{sec:liveness_detection_evaluation}).
During authentication, the test sample produces a hybrid score $S_{\mathrm{hyb}}$. Liveness is accepted if $S_{\mathrm{hyb}}$ 
falls within the decision threshold $T_L$.

\subsection{FE-Based Authentication}\label{sec:feba_scheme}

While Doppler-based liveness detection can effectively defend against spoofing, biometric systems must also address the inherent challenges of \textit{template stability} and \textit{security}. Biometric traits are noisy, non-uniform, and non-revocable, meaning that two samples from the same user are never identical. Storing raw biometric templates directly is insecure, as they cannot be replaced if compromised. 
To address these limitations, we employ \textit{FE}. An FE converts a noisy biometric measurement into a stable cryptographic key $K$ while producing a public helper $P$ to allow key reproduction. Internally, each execution samples a random secret $R$, which ensures that keys are unlinkable across multiple generations. This process leverages \textit{error-correcting codes (ECCs)} to tolerate intra-user variability and \textit{entropy extraction} to guarantee that $K$ is uniformly random.

\subsubsection{\textbf{Enrollment Phase}}\label{sec:fe_enrollment_phase}

During the enrollment phase, a cryptographic key and a public helper string are generated for each individual using inputs collected across multiple enrollment attempts. For the blow-acoustic modality, GFCC features are extracted from each recording and passed through a CNN model discussed in Section \ref{sec:gfcc_cnn}, to produce blow-acoustic embeddings. For the facial modality, facial images are processed using the FaceNet model discussed in Section \ref{sec:facenet_extraction}, to produce facial embeddings. Let $\mathrm{f}_\text{blow}^{(i)}$ and $\mathrm{f}_\text{face}^{(i)}$ denote the blow-acoustic and facial embeddings extracted from the $i$-th enrollment sample, respectively.  


\paragraph{Score-Level Fusion.} 
The blow-acoustic and facial embeddings are combined using a weighted sum 
as follows:
\begin{equation}
\mathrm{f}_\text{fused}^{(i)} = w_\text{blow} \cdot \mathrm{f}_\text{blow}^{(i)} + w_\text{face} \cdot \mathrm{f}_\text{face}^{(i)}, 
\end{equation}
where $w_\text{blow}$ and $w_\text{face}$ denotes the weighting coefficients for the blow-acoustic and facial inputs, respectively. Both coefficients are assigned equal values, i.e., $w_\text{blow} = w_\text{face} = 0.5$, as the two modalities are considered to be equally important. 

\paragraph{Feature-Level Fusion.} 
The blow-acoustic and facial embeddings are concatenated to form a single multimodal representation:
\begin{equation}
\mathrm{f}_\text{fused}^{(i)} = [\mathrm{f}_\text{blow}^{(i)}; \mathrm{f}_\text{face}^{(i)}],
\end{equation}
where ";" denotes vector concatenation. 
The resulting fused embeddings of the two techniques discussed above serve as a unified representation of both modalities and are ready for subsequent processing in the authentication pipeline.

\paragraph{Fuzzy Extractor.} 
To perform the fuzzy extraction, the fused embedding $\mathrm{f}_\text{fused}^{(i)}$ (from either fusion strategy) is first binarized using median thresholding: 
\begin{equation}
\mathrm{f}_\text{bin}^{(i)} = \text{Binarize}(\mathrm{f}_\text{fused}^{(i)}) \in \{0,1\}^M,
\end{equation}
where $M$ denotes the dimensionality of the fused embedding, which remains unchanged after binarization. 

The binarized embedding is then processed by an FE using Bose-\allowbreak Chaudhuri-Hocquenghem (BCH) ECC. During enrollment, the FE Generation function is applied to generate the cryptographic key and public helper string:
\begin{equation}
(K, P) = \text{FE.Gen}(\mathrm{f}_\text{bin}^{(i)}),
\end{equation}
where $K$ is derived from a fresh random secret $R$ (sampled internally by FE.Gen) and securely stored on the server, and $P$ is retained to allow key reproduction during authentication.


\subsubsection{\textbf{Authentication Phase}}\label{section_fe_authentication_phase}
During the authentication phase, a new blow-acoustic reading 
and facial data are captured and processed using the same feature extraction and fusion pipeline as in enrollment. The newly fused embedding $\mathrm{f}_\text{fused}'$ is first binarized to $\mathrm{f}_\text{bin}'$, which is then processed by the FE using the reproduction function:
\begin{equation}
K' = \text{FE.Rep}(\mathrm{f}_\text{bin}', P),
\end{equation}
where $P$ is the public helper string stored during enrollment. The FE leverages BCH error-correcting codes, which rely on the 
$HD$ between the query embedding $\mathrm{f}_\text{bin}'$ and the enrolled embedding $\mathrm{f}_\text{bin}$. Authentication succeeds (i.e., $K' = K$) only if 
the $HD$ falls within the acceptable threshold $T_F$, i.e., $HD \leq T_F$.
Otherwise, the cryptographic key cannot be reproduced and access is rejected.
\subsection{Key and Template Revocation}
The FE-based authentication scheme 
can be extended to support revocability. 
Specifically, since the fuzzy extractor derives both the cryptographic key and the associated public helper data from the binarized embeddings of the biometric features, new keys 
can be reissued by modifying the randomness parameters ($R$) within the fuzzy extractor 
construction. This process effectively revokes previously generated cryptographic keys and helper data in the event of compromise or whenever needed. However, biometric templates associated to physiological biometrics (e.g., fingerprint, iris, or face) cannot be revoked as these traits are permanent and non-replaceable. In contrast, behavioral biometrics offer the possibility of biometric template revocation as well since users can adopt new behavioral patterns to generate fresh 
biometric identities. Moreover, because the system relies on blow-acoustic pattern biometrics, users can also revoke the biometric template itself by adopting a new blow-acoustic pattern, generating a fresh and unique biometric identity.

In this work, since we employ blow-acoustics behavioral biometric as one modality, the biometric template itself (the behavioral part or even the combined one) can be revoked by adopting new blow-acoustic patterns. Consequently, in our framework, revocability is achieved using a fuzzy extractor 
at two levels. 
     \paragraph{1) Key Revocation.} 
    As discussed in Section \ref{sec:fe_enrollment_phase}, the fuzzy extractor allows the derivation of strong, user-specific cryptographic keys from noisy biometric inputs. In our case, the key is derived from a randomly sampled secret $R$. 
    When a compromise occurs or whenever there is a need to make changes, 
    the 
    existing key can be revoked and reissued securely using a newly sampled random secret $R'$. This dual-layer revocability tightly couples biometric authentication with cryptographic systems, offering enhanced security compared to fixed biometrics. 
   \paragraph{2) Template Revocation.} 
    As discussed in Section \ref{sec:fe_enrollment_phase}, extracted and binarized blow-acoustic and facial features are mapped into a secure representation through the fuzzy extractor. 
    Since the blow-acoustic features can be altered 
    using new blowing patterns, the system can revoke the existing biometric template and reissue a fresh one without dependency on the old or compromised template. Therefore, unlike conventional physiological biometrics (e.g., fingerprint, iris and face recognition), our biometric framework allows the revocation of both the biometric template and the associated cryptographic key. 

 This dual revocability feature distinguishes our approach from 
 conventional biometrics, 
 which lack the ability to revoke the biometric template. Therefore, by integrating revocability into our MFB framework, we address one of the fundamental limitations of conventional biometric systems: their inability to be reset or renewed even in case of compromise. This makes the proposed approach particularly well-suited for long-term deployments in security-critical applications such as smartphone unlocking, IoT device authentication, and secure access control.

\section{Experimental Evaluation}\label{sec:experimental_evaluation}

\subsection{Dataset} \label{sec:dataset}
\begin{figure*}[!h]
     \centering
     \begin{subfigure}{0.325\textwidth}
         \centering
         \includegraphics[width=\linewidth]{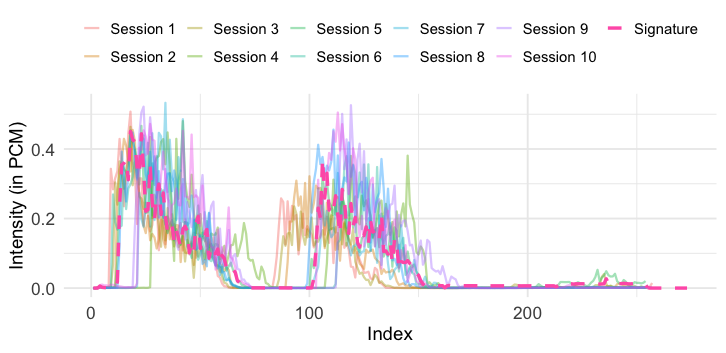}
         \caption{}
         \label{fig_participant1}
     \end{subfigure}
     \begin{subfigure}{0.325\textwidth}
         \centering
         \includegraphics[width=\linewidth]{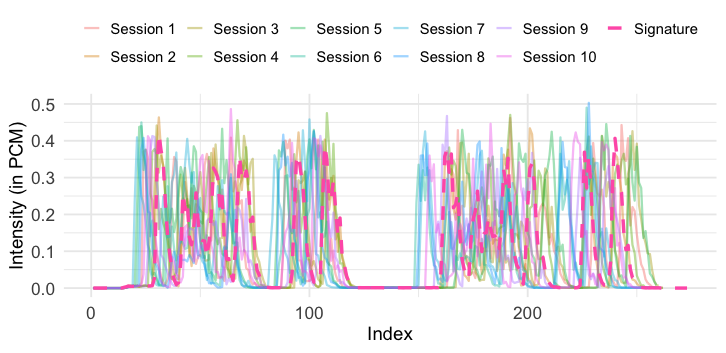}
         \caption{}
         \label{fig_participant2}
     \end{subfigure}
      \begin{subfigure}{0.325\textwidth}
         \centering
         \includegraphics[width=\linewidth]{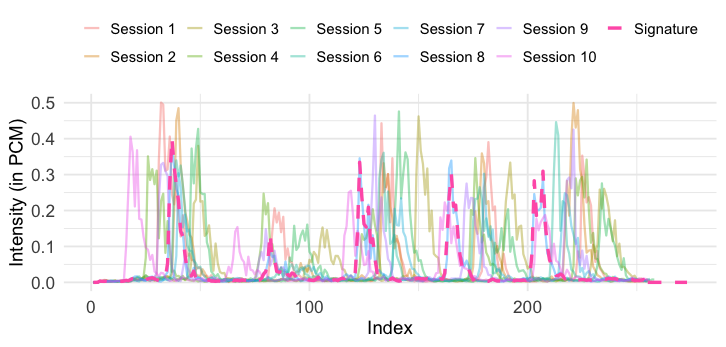}
         \caption{}
         \label{fig_participant3}
     \end{subfigure}
     \caption{Sample raw blow-acoustic data of (a) Participant 1 (b) Participant 2 (c) Participant 3.} 
     \label{fig_sample_raw_acoustic_data}
\end{figure*}

\begin{figure*}[htb]
     \centering
     \begin{subfigure}{0.325\textwidth}
         \centering
         \includegraphics[width=\linewidth]{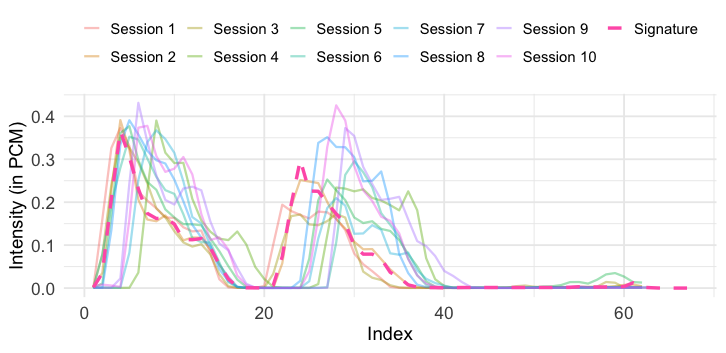}
         \caption{}
         \label{fig_participant1_r}
     \end{subfigure}
     \begin{subfigure}{0.325\textwidth}
         \centering
         \includegraphics[width=\linewidth]{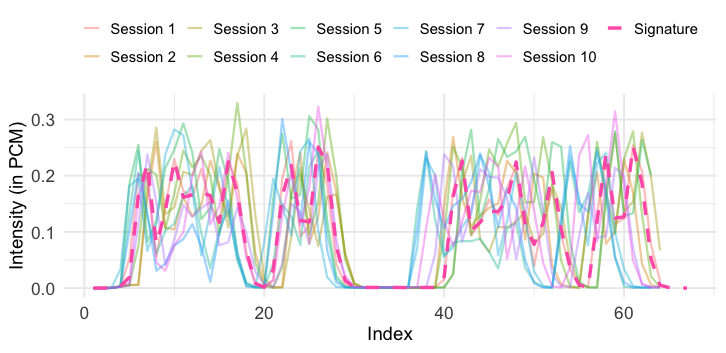}
         \caption{}
         \label{fig_participant2_r}
     \end{subfigure}
      \begin{subfigure}{0.325\textwidth}
         \centering
         \includegraphics[width=\linewidth]{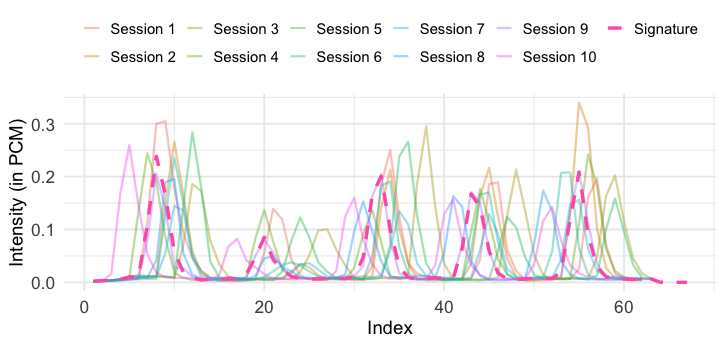}
         \caption{}
         \label{fig_participant3_3}
     \end{subfigure}
     \caption{Sample refined blow-acoustic data of (a) Participant 1 (b) Participant 2 (c) Participant 3.} 
     \label{fig_sample_refined_acoustic_data}
\end{figure*}

\begin{figure*}[!htbp]
     \centering
     \begin{subfigure}{0.325\textwidth}
         \centering
         \includegraphics[width=\linewidth]{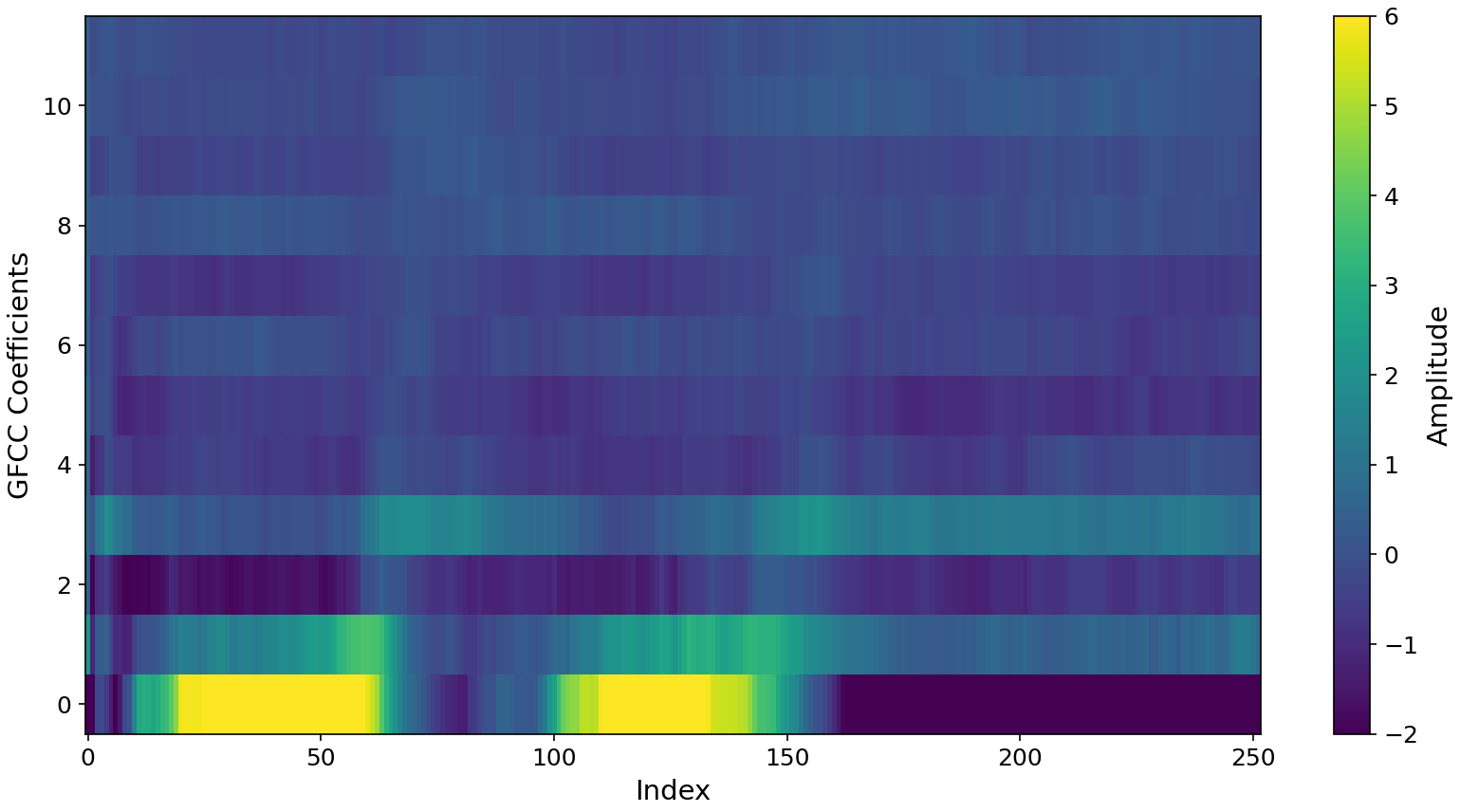}
         \caption{}
         \label{fig_gfcc_participant1}
     \end{subfigure}
     \begin{subfigure}{0.325\textwidth}
         \centering
         \includegraphics[width=\linewidth]{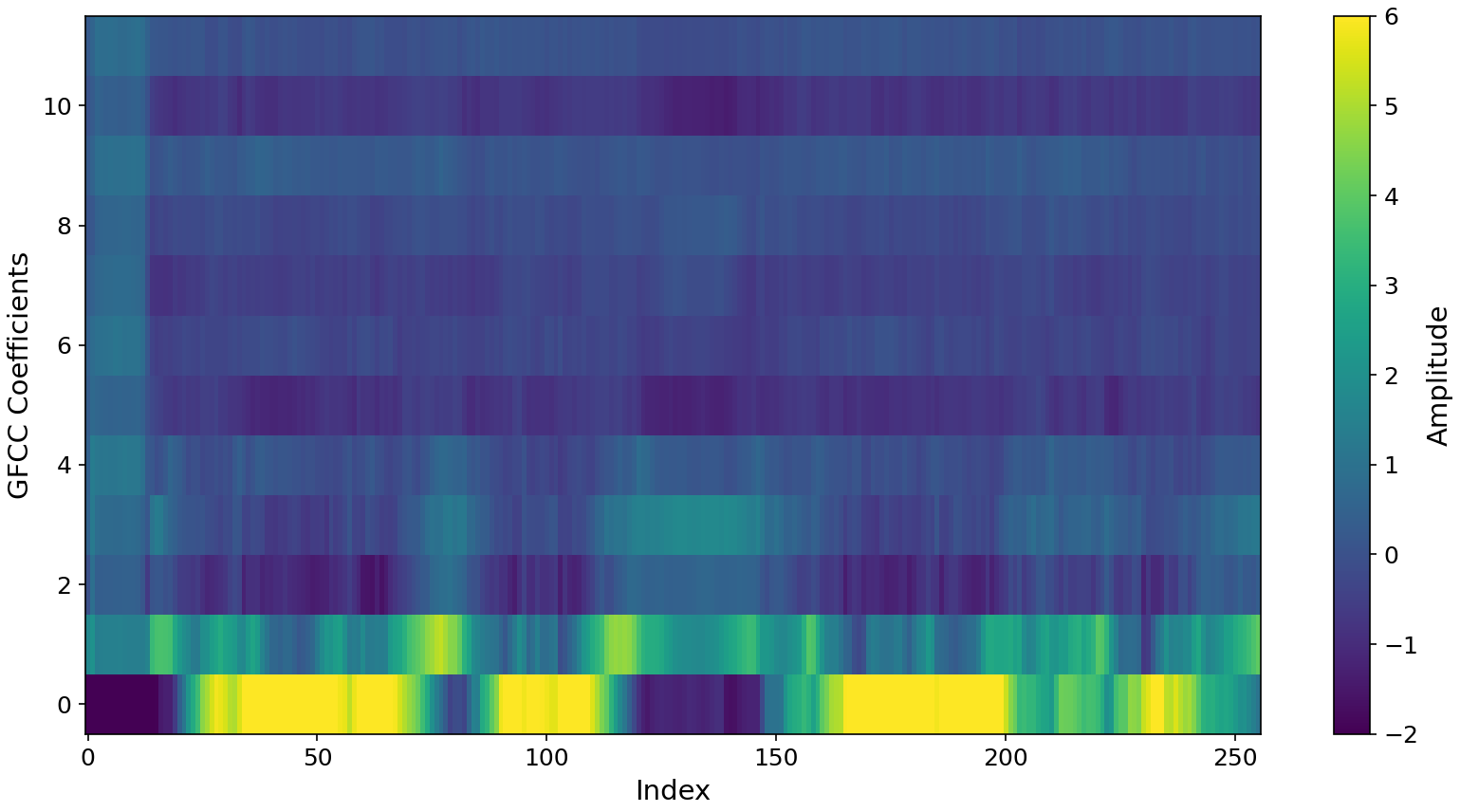}
         \caption{}
         \label{fig_gfcc_participant2}
     \end{subfigure}
      \begin{subfigure}{0.325\textwidth}
         \centering
         \includegraphics[width=\linewidth]{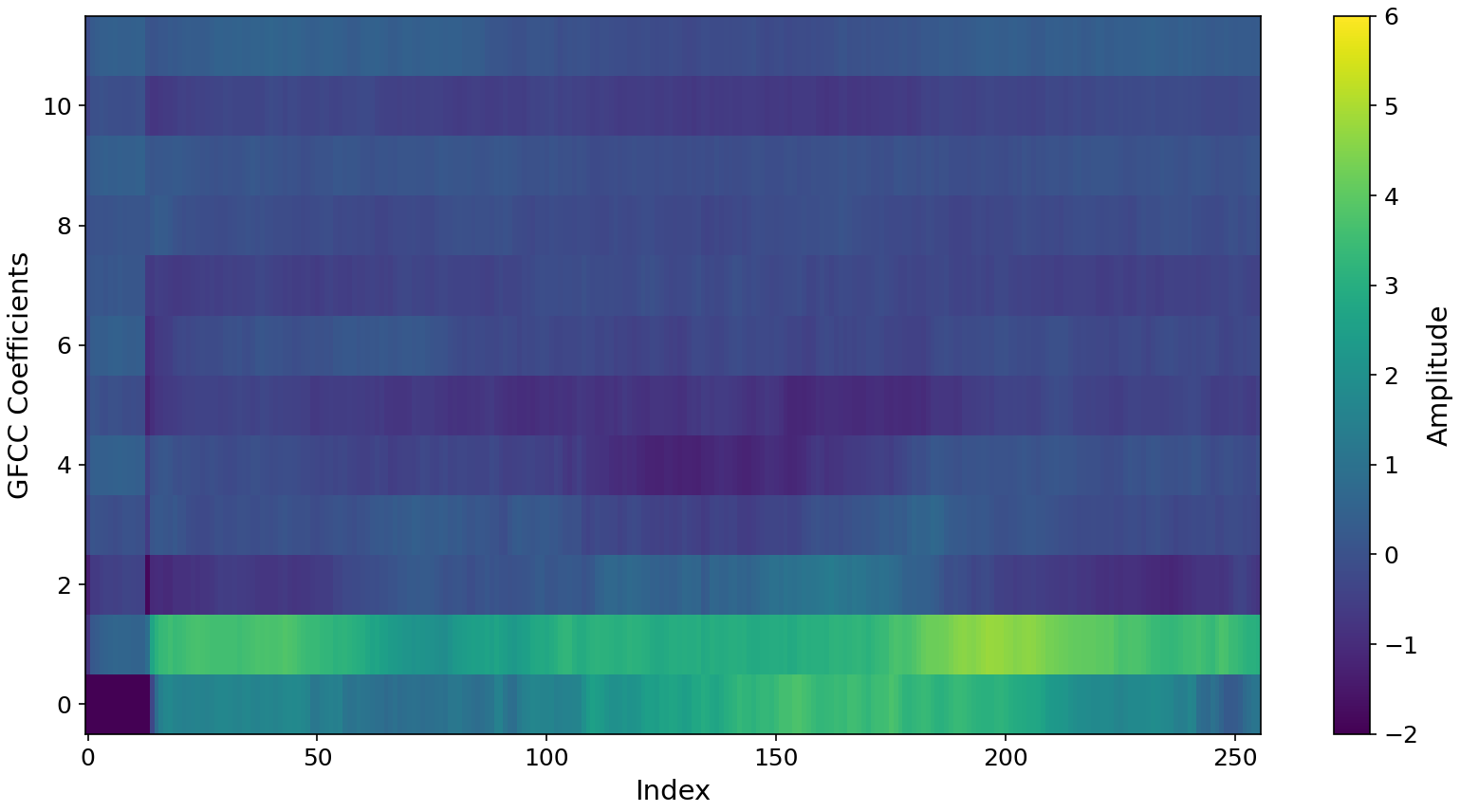}
         \caption{}
         \label{fig_gfcc_participant3}
     \end{subfigure}
     \caption{GFCC-based feature extraction for blow-acoustic data of (a) Participant 1 (b) Participant 2 (c) Participant 3}
     \label{fig_sample_gfcc_feature_extraction}
\end{figure*}

We conducted an empirical study involving \emph{50 participants} 
recruited from diverse demographic groups. 
The dataset 
was collected using our BlowLive proof-of-concept Android application described 
in Section~\ref{sec:data_acquisition}, with an ultrasonic tone emitted for the Doppler-shift based liveness detection purpose. Each participant completed 10 to 12 sessions, with 50\% in sitting and another 50\% in standing postural modes. Each session lasted approximately 5 seconds. During data collection, BlowLive captured the blow-acoustic signals using the smartphone microphone at a 48kHz sampling rate, along with facial data from the front-facing camera. The collected blow-acoustic signals and facial features were further processed using the feature extraction methods discussed in Section~\ref{sec:feature_extraction}.
To illustrate the characteristics of the collected blow-acoustic signals, Figure \ref{fig_sample_raw_acoustic_data} 
and Figure \ref{fig_sample_refined_acoustic_data} show examples of the raw and refined data, respectively, from three participants. The raw waveforms are derived by converting the captured PCM float samples into a Root Mean Square (RMS) representation to highlight amplitude variations over time. 
It demonstrates the distinct blowing patterns exhibited by different users across 10 sessions of 5 seconds each. The red dotted “Signature” line represents an aggregated trajectory computed using the DBA algorithm\cite{Petitjean2011_DBA}. These two representations are primarily used to visualize the blow patterns and are not directly used as features for enrollment and authentication. 
Figure \ref{fig_sample_gfcc_feature_extraction} illustrates samples of the GFCC-extracted features discussed in Section \ref{sec:GFCC_feature_extraction}. Although the full GFCC feature set consists of 26 static coefficients along with their delta components, the figure displays only the first 13 cepstral coefficients for clarity.
The datasets used for FE-based authentication and liveness detection, along with the corresponding implementation and reproducibility instructions, are publicly available in \cite{blowlive_dataset}, specifically under the ``\texttt{Authentication}'' and ``\texttt{LivenessDetection}'' folders, respectively. The latter dataset involves blow-acoustic signals with emitted ultrasonic tones for the Doppler-shift based liveness detection analysis.


 \subsection{Liveness Detection Evaluation}\label{sec:liveness_detection_evaluation}

\begin{figure}[!htbp]
     \centering
     \begin{subfigure}{0.48\textwidth}
         \centering
         \includegraphics[width=\linewidth]{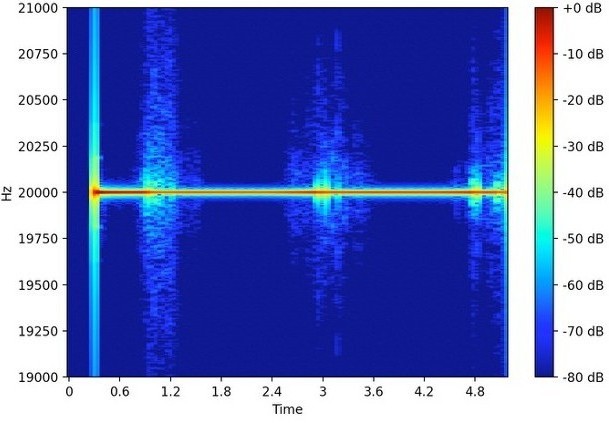}
         \caption{}
         \label{fig:doppler_live_blow}
     \end{subfigure}
     \begin{subfigure}{0.48\textwidth}
         \centering
         \includegraphics[width=\linewidth]{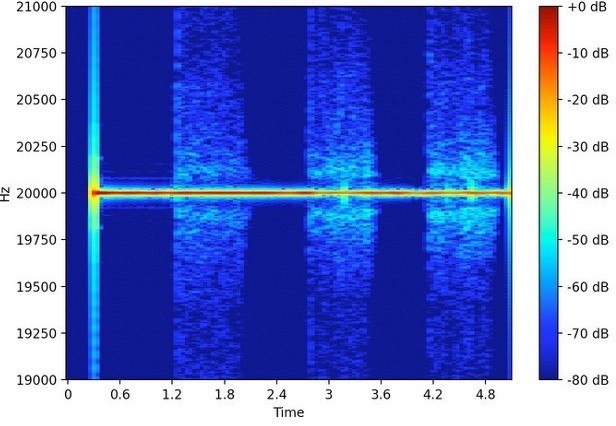}
         \caption{}
         \label{fig:doppler_playback_attack}
     \end{subfigure}
     \caption{Doppler shifts for (a) live blow, (b) playback acoustic signals}
     \label{fig:doppler_shift_samples}
\end{figure}

To evaluate 
the proposed Doppler shift-based liveness detection module, we collected 554 legitimate and 558 playback biometric data from 50 participants (the dataset and implementation is provided in \cite{blowlive_dataset}, under the ``\texttt{LivenessDetection}'' folder). The Doppler shift characteristics of legitimate (live blow) and playback acoustics signals are demonstrated in Figure \ref{fig:doppler_live_blow} and \ref{fig:doppler_playback_attack}, respectively. As shown in Figure \ref{fig:doppler_live_blow}, prominent frequency cues are observed around the $20~\mathrm{kHz}$ ultrasonic band, clearly indicating the presence of Doppler shifts due to live blowing acoustic signals. In contrast, such cues are largely absent in Figure \ref{fig:doppler_playback_attack}, demonstrating the lack of Doppler shift in playback signals.  

Ultimately, we evaluate accuracy of the module using global and dynamic (per-user) threshold settings (see Section \ref{sec:threshold_computation} for details). 
The experimental results are depicted in Table \ref{tab:liveness_results}. In the global setting, we achieve an accuracy of $90.38\%$, with an FAR (false acceptance rate) of $5.56\%$ and an FRR (false rejection rate) of $13.72\%$.  
In contrast, the per-user setting yields substantially higher performance, achieving $99.46\%$ accuracy with a FAR of $0\%$ and an FRR of $1.08\%$. 
The superior performance of the per-user setting is expected, as airflow strength and blowing characteristics vary significantly across individuals. 

\begin{table}[htb]
\centering
\caption{Doppler shift-based liveness detection accuracy}
\label{tab:liveness_results}
\begin{tabular}{lccc}
\toprule
\textbf{Method} & \textbf{Accuracy (\%)} & \textbf{FAR (\%)} & \textbf{FRR (\%)} 
 \\
\midrule
Global Threshold & 90.38 & 5.56 & 13.72 
\\
Dynamic Threshold (Per-User) & 99.46 & 0 & 1.08 
\\
\bottomrule
\end{tabular}
\end{table}

\subsection{FE-Based Authentication Evaluation} \label{sec:fe_accuracy_evaluation}

\begin{table*}[htb]
\caption{Accuracy (in \%) of the 
FE-based authentication for a variety of K-NN values over various target recall values $q$ and a dynamic threshold $T_F$. 
For each $q$, the highest accuracy and lowest 
FAR values in the ``Both'' mode are marked in bold.}
\label{tab:feba_results}
\centering
\begin{tabular}{cl|ccc|ccc|ccc}
\multirow{2}{*}{K-NN} &
  \multicolumn{1}{c|}{\multirow{2}{*}{Mode}} &
  \multicolumn{3}{c|}{\begin{tabular}[c]{@{}c@{}}q = 5 (Sit \& Stand)\\ q = 10 (Both)\end{tabular}} &
  \multicolumn{3}{c|}{\begin{tabular}[c]{@{}c@{}}q = 4 (Sit \& Stand)\\ q = 9 (Both)\end{tabular}} &
  \multicolumn{3}{c}{q = 8 (Both)} \\
                   & \multicolumn{1}{c|}{} &
                   accuracy & FAR    & FRR & accuracy & FAR    & FRR   & accuracy & FAR    & FRR   \\ \hline
\multirow{3}{*}{1} & Sit                   & 
                    99.29   & 0.73 & 0   & 99.54   & 0.13 & 16  & -        & -      & -     \\
                   & Stand                 & 
                   99.64   & 0.37 & 0   & 99.57   & 0.16 & 13.6 & -        & -      & -     \\
                   & Both                  & 
                   \textbf{99.09}& \textbf{0.93} & 0   & \textbf{99.56}   & \textbf{0.30} & 7.6 & \textbf{99.56}   & \textbf{0.15} & 14.4 \\ \hline
\multirow{3}{*}{2} & Sit                   & 
                    99.30   & 0.72 & 0   & 99.56   & 0.11 & 16.8 & -        & -      & -     \\
                   & Stand                 & 
                   99.43   & 0.58 & 0   & 99.53   & 0.20 & 13.6 & -        & -      & -     \\
                   & Both                  & 
                   98.79   & 1.24 & 0   & 99.52   & 0.34 & 7.4 & 99.49   & 0.21 & 15  \\ \hline
\multirow{3}{*}{3} & Sit                   & 
                    99.16   & 0.86 & 0   & 99.55   & 0.14 & 15.6 & -        & -      & -     \\
                   & Stand                 & 
                   99.46   & 0.55 & 0   & 99.43   & 0.25 & 16  & -        & -      & -     \\
                   & Both                  & 
                   98.88   & 1.14 & 0   & 99.56   & 0.31 & 7  & 99.54   & 0.18 & 14.2 \\ \hline
\multirow{3}{*}{4} & Sit                   & 
                    98.93   & 1.09 & 0   & 98.93   & 1.09 & 0     & -        & -      & -     \\
                   & Stand                 & 
                   99.42   & 0.59 & 0   & 99.42   & 0.59 & 0     & -        & -      & -     \\
                   & Both                  & 
                   98.94   & 1.09 & 0   & 99.42   & 0.44 & 7.4 & 99.52   & 0.19 & 14.8
\end{tabular}%
\end{table*}

\begin{table*}[htb]
\caption{Accuracy (in \%)) of the FE-based authentication using the Doppler-based dataset} 
\label{tab:doppler_based_auth_accuracy}
\centering
\begin{tabular}{c|ccc|ccc|ccc}
\multirow{2}{*}{K-NN} & 
\multicolumn{3}{c|}{q = 10}        &
\multicolumn{3}{c|}{q = 9}            & \multicolumn{3}{c}{q = 8}              \\
  & accuracy & FAR  & FRR & accuracy & FAR  & FRR  & accuracy & FAR  & FRR   \\ \hline
1  &
\textbf{96.58} & \textbf{3.49} & 0 & \textbf{99.20} & \textbf{0.64} & 8.96 & \textbf{99.42} & \textbf{0.24} & 17.26 \\ \hline
2 & 
95.63    & 4.46 & 0   & 98.65    & 1.20 & 8.83 & 99.13    & 0.51 & 18.32 \\ \hline
3 & 
94.45    & 5.66 & 0   & 98.28    & 1.57 & 8.79 & 98.84    & 0.81 & 18.32 \\ \hline
4 & 
92.28    & 7.88 & 0   & 97.56    & 2.30 & 8.96 & 98.28    & 1.39 & 17.80
\end{tabular}%
\end{table*}


In this section, we evaluate the overall accuracy of our FE-based authentication scheme discussed in Section \ref{sec:feba_scheme} using the dataset collected from 50 participants (the dataset and implementation is provided in \cite{blowlive_dataset}, under the ``\texttt{Authentication}'' folder). 
We evaluate three different scenarios: the sitting dataset, the standing dataset, and the combination of the two. Moreover, 
multiple $k$-NN values  (i.e., $k$) and target recall values ($q$) are considered to compute the evaluation metrics. 


Table \ref{tab:feba_results} displays results of 
the 
accuracy, FAR, and FRR 
using HD as the similarity metrics. The evaluation is conducted under the three scenarios, parameters ($k$ and $q$), and threshold ($T_F$) settings used in Table \ref{tab:feba_results}. The highest accuracy (99.56\%) and the lowest FAR (0.15\%) 
are achieved at $k$=1 and $q$=8 (with the same accuracy also obtained at $q=9$). The lowest FRR (0\%) is achieved at $q=10$ for all $k$ values. 
Notably, varying the $k$ value results in only minor fluctuations across all metrics, a trend consistent with the behavior observed in the blow-acoustic evaluation.

We also evaluate the authentication accuracy of the proposed technique on the Doppler shift-based dataset, which incorporates blow-acoustic signals with emitted ultrasonic tones that was intended to be used for the liveness detection. The result is depicted in Table \ref{tab:doppler_based_auth_accuracy}. 
While some instabilities are observed across different $k$ and $q$ values, presumably due to the emitted ultrasonic tones, the accuracy is still considerably high, such as 99.42\% at $k = 1$ and $q = 8$.  

Table \ref{tab:revocability_combined_results} summarizes the 
score-level and feature-level fusion accuracy results of the blow-acoustic and facial-recognition modalities. In particular, the score-level fusion achieves 99.95\% accuracy and 0\% FRR 
at $q=10$, while it achieves 0\% FAR at $q=9$. On the other hand, the feature-level fusion achieves 100\% accuracy, with 0\% of FAR and FRR 
at $q=10$. 
Notably, the score-level fusion exhibits slightly inferior performance. Although it still maintains a FAR of 0\% for most thresholds, its FRR and accuracy values degrade more noticeably as $q$ decreases. This suggests that aggregating similarity scores is less effective because it cannot capture complementary information from both modalities. 
In contrast, feature-level fusion achieves the best overall performance across all $q$ values. 


\begin{table*}[htb]
\caption{Accuracy (in \%)) of the FE-based authentication when blow-acoustic and face recognition outputs are fused via score-level and feature-level fusion.} 
\label{tab:revocability_combined_results}
\centering
\begin{tabular}{c|ccc|ccc|ccc}
\multirow{2}{*}{\begin{tabular}[c]{@{}c@{}}Biometrics\\ Features\end{tabular}} &
  \multicolumn{3}{c|}{q = 10} &
  \multicolumn{3}{c|}{q = 9} &
  \multicolumn{3}{c}{q = 8} \\
                     &  
                     accuracy & FAR    & FRR   & accuracy & FAR    & FRR   & accuracy & FAR    & FRR   \\ \hline
Blow-Acoustic        & 
99.09   & 0.93 & 0 & 99.56   & 0.30 & 7.6 & 99.56   & 0.15 & 14.4 \\ \hline
Facial-Recognition   & 
100   & 0 & 0 & 99.88   & 0 & 6 & 99.74   & 0 & 12.8 \\ \hline
Score-Level Fusion   & 
\textbf{99.95}   & 0.05 & \textbf{0} & 99.81   & \textbf{0} & 9.4 & 99.66   & 0 & 16.8 \\ \hline
Feature-Level Fusion & 
\textbf{100}   & \textbf{0} & \textbf{0} & 99.89   & 0 & 5.4 & 99.76   & 0 & 12 \\ \hline
\end{tabular}%
\end{table*}

\subsection{\textbf{Revocability Evaluation}}
From a security perspective, the FE scheme exhibits several essential properties that ensure robust and privacy-preserving authentication. 

\textbf{Unlinkability}: each cryptographic key $K$ is derived from a freshly sampled random secret $R$ and associated helper data $P$, such that $K$ does not reveal information about the underlying biometric embedding. This ensures that the same biometric cannot be used to link keys across different applications or enrollment instances, preventing cross-system tracking and preserving privacy.

\textbf{Non-invertibility}: the combination of the random secret $R$ and the BCH-based helper data $P$ does not reveal sufficient information to reconstruct the original biometric embedding, protecting the template from inversion attacks.

\textbf{Diversity}: the use of a randomly sampled secret $R$ ensures that the cryptographic keys $K$ occupy a large, diverse key space, so that each key derived from the same biometric embedding is effectively independent and unpredictable.

\textbf{Revocability}: if a key is compromised or requires renewal, a new random secret $R'$ can be sampled to produce a fresh key and helper data without altering the underlying biometric template, or a new set of  blow-acoustic behavioral biometric patterns 
can be introduced. 
This enables secure revocation and reissuance of both the cryptographic key and biometric template in our scheme.

Together, these properties ensure that the FE scheme provides a secure, unlinkable, and revocable cryptographic key while maintaining the privacy of the underlying biometric features.


\subsection{Evaluation of Other Requirements}


BlowLive inherently achieves other requirements, including resilience against known attacks, non-invasiveness, usability, MFB support and low resource requirements. 

\textbf{Resilience Against Known Attacks}: As discussed in previous sections, BlowLive is explicitly designed to defend against a broad spectrum of attacks that commonly compromise biometric authentication systems. The incorporated multi-factor authentication techniques and the introduced liveness detection mechanism enable BlowLive to remain resilient against a wide range of spoofing, presentation, template leakage, and other adversarial attack techniques.

\textbf{Non-Invasiveness}: BlowLive is is entirely contactless and touch-free, requiring only a simple blow directed toward the phone screen. The procedure is completed within a matter of seconds, making it highly non-invasive and significantly reducing hygiene-related concerns as well as the risk of privacy violations associated with physical contact.

\textbf{Usability}: BlowLive is user-friendly and convenient for authentication, requiring only a simple blow toward the phone along with a simultaneous facial snapshot to complete the process. As such, it constitutes an intuitive, non-intrusive, and highly practical multi-factor biometric modality.

\textbf{MFB Support}: BlowLive seamlesssly integrates blow-acoustic and facial biometric modalities, achieving multi-factor biometrics with high robustness and accuracy. 

\textbf{Low Resource Requirements}: BlowLive does not require any additional or specialized hardware or software and is compatible with low-specification smartphones equipped only with a camera and a microphone.

\subsection{Performance Evaluation}

Although the proposed authentication framework is tested offline, we measured and evaluated the runtime performance of the main operations involved in the authentication pipeline. The runtime performance was measured on a MacBook Pro equipped with a 2.4 GHz quad-core i5 CPU, 16 GB RAM, and an Iris Plus integrated GPU. The system operates strictly in authentication (1:1) mode rather than identification (1:N). Consequently, the computational complexity per request is independent of the total number of enrolled users. The detailed execution time and time complexity of 
the main functions in BlowLive are summarized in Table~\ref{tab:performance_evaluation} 
Since only a single 
fixed-size helper data 
is stored per user for the FE-based authentication, the total server-side storage scales linearly with the number of enrolled users. For the user-side, the storage requirement is 
negligible. 


\begin{table}[htb]
\centering
\caption{Performance evaluation ($N$: 
biometric signal length; $M$: dimensional of the embeddings length 
}
\label{tab:performance_evaluation}
\begin{tabular}{lllcc}
\hline
\textbf{Method}           & \multicolumn{2}{l}{\textbf{Function}}   & \textbf{Time (ms)} & \textbf{Complexity}\\ \hline
\multirow{2}{*}{\makecell[l]{Feature \\Extraction}} & \multicolumn{2}{l}{GFCC Feature Extraction}   & 217.205 & $O(N)$            \\
                          & \multicolumn{2}{l}{CNN Inference} & 2.3750  & $O(N)$           \\
\multirow{3}{*}{\makecell[l]{Authentication \& \\Revocation}} & \multicolumn{2}{l}{FE Key Generation}   & 0.0224 & $O(M)$            \\
                          & \multicolumn{2}{l}{FE Key Reproduction} & 1.0667  & $O(M)$           \\
                          & \multicolumn{2}{l}{FE Key Revocation}       & 0.1304    & $O(M)$         \\
Liveness Detection                    & \multicolumn{2}{l}{Doppler Shift Analysis} & 
234.956 & $O(N)$ \\ \hline
\end{tabular}%
\vspace{-0.6cm}
\end{table}

\subsection{Comparison with Related Works}

In Table~\ref{table_comp_related_works}, we summarize the comparison between BlowLive and the most closely related work discussed in Section~\ref{sec:related_work}. For each work, we present the best reported accuracy in terms of the metrics discussed in this section. Furthermore, 
we compare each work based on their reported best accuracy and whether they meet the other requirements discussed in Section~\ref{sec:requirements}. 
We observe that BlowLive achieves the highest accuracy and it is 
the only work to satisfy all requirements.

\begin{table*}[htb]

\centering
\caption{Qualitative and quantitative comparison of related works.} 
\label{table_comp_related_works}
\resizebox{\textwidth}{!}{%
\begin{tabular}{|lcl|ccccccccccc|}
\hline
\multicolumn{3}{|l|}{
\multirow{3}{*}{Related work}} &
  \multicolumn{11}{c|}{Requirements} \\ \cline{4-14} 
\multicolumn{3}{|l|}{} &
  \multicolumn{3}{c|}{R1} &
  \multicolumn{1}{c|}{\multirow{2}{*}{R2}} &
  \multicolumn{2}{c|}{R3} &
  \multicolumn{1}{c|}{\multirow{2}{*}{R4}} &
  \multicolumn{1}{c|}{\multirow{2}{*}{R5}} &
  \multicolumn{1}{c|}{\multirow{2}{*}{R6}} &
  \multicolumn{1}{c|}{\multirow{2}{*}{R7}} &
  \multirow{2}{*}{R8} \\ \cline{4-6} \cline{8-9}
\multicolumn{3}{|l|}{} &
  \multicolumn{1}{c|}{Acc.} &
  \multicolumn{1}{c|}{FAR} &
  \multicolumn{1}{c|}{FRR} &
  \multicolumn{1}{c|}{} &
  \multicolumn{1}{c|}{KR} &
  \multicolumn{1}{c|}{TR} &
  \multicolumn{1}{c|}{} &
  \multicolumn{1}{c|}{} &
  \multicolumn{1}{c|}{} &
  \multicolumn{1}{c|}{} &
   \\ \hline
\multicolumn{3}{|l|}{Chauhan et al. \cite{chauhan2017breathprint}} &
  \multicolumn{1}{c|}{—} &
  \multicolumn{1}{c|}{$2\%$} &
  \multicolumn{1}{c|}{$6\%$*} &
  \multicolumn{1}{c|}{$\times$} &
  \multicolumn{1}{c|}{$\times$} &
  \multicolumn{1}{c|}{$\times$} &
  \multicolumn{1}{c|}{$\checkmark$} &
  \multicolumn{1}{c|}{$\checkmark$} &
  \multicolumn{1}{c|}{$\checkmark$} &
  \multicolumn{1}{c|}{$\times$} &
  $\checkmark$ \\ \hline
\multicolumn{3}{|l|}{Al-Waisy et al. \cite{al2017multimodal}} &
  \multicolumn{1}{c|}{99\%} &
  \multicolumn{1}{c|}{---} &
  \multicolumn{1}{c|}{---} &
  \multicolumn{1}{c|}{$\times$} &
  \multicolumn{1}{c|}{$\checkmark$} &
  \multicolumn{1}{c|}{$\checkmark$} &
  \multicolumn{1}{c|}{$\checkmark$} &
  \multicolumn{1}{c|}{$\checkmark$} &
  \multicolumn{1}{c|}{$\times$} &
  \multicolumn{1}{c|}{$\times$} &
  $\times$ \\ \hline
\multicolumn{3}{|l|}{Aizi et al. \cite{aizi2022score}} &
  \multicolumn{1}{c|}{95\%} &
  \multicolumn{1}{c|}{3.89\%} &
  \multicolumn{1}{c|}{1.5\%} &
  \multicolumn{1}{c|}{$\times$} &
  \multicolumn{1}{c|}{$\times$} &
  \multicolumn{1}{c|}{$\times$} &
  \multicolumn{1}{c|}{$\times$} &
  \multicolumn{1}{c|}{$\times$} &
  \multicolumn{1}{c|}{$\checkmark$} &
  \multicolumn{1}{c|}{$\checkmark$} &
  $\checkmark$ \\ \hline
\multicolumn{3}{|l|}{Srivastava et al. \cite{srivastava2022match}} &
  \multicolumn{1}{c|}{99.7\%} &
  \multicolumn{1}{c|}{—} &
  \multicolumn{1}{c|}{—} &
  \multicolumn{1}{c|}{$\times$} &
  \multicolumn{1}{c|}{$\times$} &
  \multicolumn{1}{c|}{$\times$} &
  \multicolumn{1}{c|}{$\times$} &
  \multicolumn{1}{c|}{$\times$} &
  \multicolumn{1}{c|}{$\times$} &
  \multicolumn{1}{c|}{$\checkmark$} &
  $\times$ \\ \hline
\multicolumn{3}{|l|}{Mahfouz et al. \cite{mahfouz2024m2auth}} &
  \multicolumn{1}{c|}{—} &
  \multicolumn{1}{c|}{—} &
  \multicolumn{1}{c|}{—} &
  \multicolumn{1}{c|}{$\times$} &
  \multicolumn{1}{c|}{$\times$} &
  \multicolumn{1}{c|}{$\times$} &
  \multicolumn{1}{c|}{$\times$} &
  \multicolumn{1}{c|}{$\times$} &
  \multicolumn{1}{c|}{$\checkmark$} &
  \multicolumn{1}{c|}{$\checkmark$} &
  $\checkmark$ \\ \hline
\multicolumn{3}{|l|}{El Rahman et al. \cite{rahman2020multimodal}} &
  \multicolumn{1}{c|}{---} &
  \multicolumn{1}{c|}{---} &
  \multicolumn{1}{c|}{---} &
  \multicolumn{1}{c|}{$\checkmark$} &
  \multicolumn{1}{c|}{$\checkmark$} &
  \multicolumn{1}{c|}{$\times$} &
  \multicolumn{1}{c|}{$\checkmark$} &
  \multicolumn{1}{c|}{$\times$} &
  \multicolumn{1}{c|}{$\times$} &
  \multicolumn{1}{c|}{$\times$} &
  $\times$ \\ \hline
\multicolumn{3}{|l|}{Lee et al. \cite{lee2021advanced}} &
  \multicolumn{1}{c|}{83\%} &
  \multicolumn{1}{c|}{1 -- 7\%} &
  \multicolumn{1}{c|}{—} &
  \multicolumn{1}{c|}{$\times$} &
  \multicolumn{1}{c|}{$\times$} &
  \multicolumn{1}{c|}{$\times$} &
  \multicolumn{1}{c|}{$\checkmark$} &
  \multicolumn{1}{c|}{$\times$} &
  \multicolumn{1}{c|}{$\checkmark$} &
  \multicolumn{1}{c|}{$\checkmark$} &
  $\times$ \\ \hline
\multicolumn{3}{|l|}{Wu et al. \cite{wu2022echohand}} &
  \multicolumn{1}{c|}{—} &
  \multicolumn{1}{c|}{—} &
  \multicolumn{1}{c|}{—} &
  \multicolumn{1}{c|}{$\times$} &
  \multicolumn{1}{c|}{$\times$} &
  \multicolumn{1}{c|}{$\times$} &
  \multicolumn{1}{c|}{$\checkmark$} &
  \multicolumn{1}{c|}{$\checkmark$} &
  \multicolumn{1}{c|}{$\checkmark$} &
  \multicolumn{1}{c|}{$\checkmark$} &
  $\checkmark$ \\ \hline
\multicolumn{3}{|l|}{Zhou et al. \cite{zhou2018echoprint}} &
  \multicolumn{1}{c|}{93.75\%} &
  \multicolumn{1}{c|}{—} &
  \multicolumn{1}{c|}{10\%} &
  \multicolumn{1}{c|}{$\times$} &
  \multicolumn{1}{c|}{$\times$} &
  \multicolumn{1}{c|}{$\times$} &
  \multicolumn{1}{c|}{$\checkmark$} &
  \multicolumn{1}{c|}{$\checkmark$} &
  \multicolumn{1}{c|}{$\checkmark$} &
  \multicolumn{1}{c|}{$\checkmark$} &
  $\checkmark$ \\ \hline
\multicolumn{3}{|l|}{Lu et al. \cite{lu2018lippass}} &
  \multicolumn{1}{c|}{90.21 -- 93.1\%} &
  \multicolumn{1}{c|}{—} &
  \multicolumn{1}{c|}{—} &
  \multicolumn{1}{c|}{$\checkmark$} &
  \multicolumn{1}{c|}{$\times$} &
  \multicolumn{1}{c|}{$\times$} &
  \multicolumn{1}{c|}{$\checkmark$} &
  \multicolumn{1}{c|}{$\checkmark$} &
  \multicolumn{1}{c|}{$\checkmark$} &
  \multicolumn{1}{c|}{$\times$} &
  $\checkmark$ \\ \hline
\multicolumn{3}{|l|}{Wu et al. \cite{wu2019lvid}} &
  \multicolumn{1}{c|}{93.47 -- 95\%} &
  \multicolumn{1}{c|}{—} &
  \multicolumn{1}{c|}{—} &
  \multicolumn{1}{c|}{$\checkmark$} &
  \multicolumn{1}{c|}{$\times$} &
  \multicolumn{1}{c|}{$\times$} &
  \multicolumn{1}{c|}{$\checkmark$} &
  \multicolumn{1}{c|}{$\checkmark$} &
  \multicolumn{1}{c|}{$\checkmark$} &
  \multicolumn{1}{c|}{$\checkmark$} &
  $\checkmark$ \\ \hline
\multicolumn{3}{|l|}{Zhang et al. \cite{zhang2022continuous}} &
  \multicolumn{1}{c|}{99\%} &
  \multicolumn{1}{c|}{—} &
  \multicolumn{1}{c|}{—} &
  \multicolumn{1}{c|}{$\checkmark$} &
  \multicolumn{1}{c|}{$\times$} &
  \multicolumn{1}{c|}{$\times$} &
  \multicolumn{1}{c|}{$\checkmark$} &
  \multicolumn{1}{c|}{$\checkmark$} &
  \multicolumn{1}{c|}{$\checkmark$} &
  \multicolumn{1}{c|}{$\checkmark$} &
  $\checkmark$ \\ \hline
\multicolumn{3}{|l|}{Zhou et al. \cite{zhou2024securing}} &
  \multicolumn{1}{c|}{96.6\%} &
  \multicolumn{1}{c|}{---} &
  \multicolumn{1}{c|}{---} &
  \multicolumn{1}{c|}{$\checkmark$} &
  \multicolumn{1}{c|}{$\times$} &
  \multicolumn{1}{c|}{$\times$} &
  \multicolumn{1}{c|}{$\checkmark$} &
  \multicolumn{1}{c|}{$\checkmark$} &
  \multicolumn{1}{c|}{$\checkmark$} &
  \multicolumn{1}{c|}{$\checkmark$} &
  $\checkmark$ \\ \hline
\multicolumn{3}{|l|}{Liu et al. \cite{liu2022contactless}} &
  \multicolumn{1}{c|}{74-98\%} &
  \multicolumn{1}{c|}{---} &
  \multicolumn{1}{c|}{---} &
  \multicolumn{1}{c|}{$\checkmark$} &
  \multicolumn{1}{c|}{$\times$} &
  \multicolumn{1}{c|}{$\times$} &
  \multicolumn{1}{c|}{$\times$} &
  \multicolumn{1}{c|}{$\checkmark$} &
  \multicolumn{1}{c|}{$\checkmark$} &
  \multicolumn{1}{c|}{$\times$} &
  $\checkmark$ \\ \hline
\multicolumn{3}{|l|}{Yang et al. \cite{yang2022linear}} &
  \multicolumn{1}{c|}{---} &
  \multicolumn{1}{c|}{---} &
  \multicolumn{1}{c|}{---} &
  \multicolumn{1}{c|}{$\times$} &
  \multicolumn{1}{c|}{$\checkmark$} &
  \multicolumn{1}{c|}{$\times$} &
  \multicolumn{1}{c|}{$\checkmark$} &
  \multicolumn{1}{c|}{$\times$} &
  \multicolumn{1}{c|}{$\checkmark$} &
  \multicolumn{1}{c|}{$\times$} &
  $\checkmark$ \\ \hline
\multicolumn{3}{|l|}{Hmani et al. \cite{hmani2024revocable}} &
  \multicolumn{1}{c|}{---} &
  \multicolumn{1}{c|}{0\%} &
  \multicolumn{1}{c|}{0.8\%} &
  \multicolumn{1}{c|}{$\times$} &
  \multicolumn{1}{c|}{$\checkmark$} &
  \multicolumn{1}{c|}{$\times$} &
  \multicolumn{1}{c|}{$\checkmark$} &
  \multicolumn{1}{c|}{$\checkmark$} &
  \multicolumn{1}{c|}{$\checkmark$} &
  \multicolumn{1}{c|}{$\checkmark$} &
  $\checkmark$ \\ \hline
\multicolumn{3}{|l|}{Imran et al. \cite{imran2025privacy}} &
  \multicolumn{1}{c|}{---} &
  \multicolumn{1}{c|}{<1\%} &
  \multicolumn{1}{c|}{1\%} &
  \multicolumn{1}{c|}{$\times$} &
  \multicolumn{1}{c|}{$\checkmark$} &
  \multicolumn{1}{c|}{$\times$} &
  \multicolumn{1}{c|}{$\checkmark$} &
  \multicolumn{1}{c|}{$\times$} &
  \multicolumn{1}{c|}{$\checkmark$} &
  \multicolumn{1}{c|}{$\times$} &
  $\checkmark$ \\ \hline
\multicolumn{3}{|l|}{Champaneria et al. \cite{champaneria2025cancelable}} &
  \multicolumn{1}{c|}{---} &
  \multicolumn{1}{c|}{---} &
  \multicolumn{1}{c|}{---} &
  \multicolumn{1}{c|}{$\times$} &
  \multicolumn{1}{c|}{$\checkmark$} &
  \multicolumn{1}{c|}{$\times$} &
  \multicolumn{1}{c|}{$\checkmark$} &
  \multicolumn{1}{c|}{$\times$} &
  \multicolumn{1}{c|}{$\checkmark$} &
  \multicolumn{1}{c|}{$\times$} &
  $\times$ \\ \hline
\multicolumn{3}{|l|}{Belhocine et al. \cite{belhocine2025medbioch}} &
  \multicolumn{1}{c|}{---} &
  \multicolumn{1}{c|}{0.00013-0.073\%} &
  \multicolumn{1}{c|}{0-0.048\%} &
  \multicolumn{1}{c|}{$\checkmark$} &
  \multicolumn{1}{c|}{$\checkmark$} &
  \multicolumn{1}{c|}{$\times$} &
  \multicolumn{1}{c|}{$\times$} &
  \multicolumn{1}{c|}{$\times$} &
  \multicolumn{1}{c|}{$\times$} &
  \multicolumn{1}{c|}{$\checkmark$} &
  $\times$ \\ \hline
\multicolumn{3}{|l|}{Ragendhu et al. \cite{ragendhu2024cancelable}} &
  \multicolumn{1}{c|}{---} &
  \multicolumn{1}{c|}{---} &
  \multicolumn{1}{c|}{---} &
  \multicolumn{1}{c|}{$\times$} &
  \multicolumn{1}{c|}{$\checkmark$} &
  \multicolumn{1}{c|}{$\times$} &
  \multicolumn{1}{c|}{$\checkmark$} &
  \multicolumn{1}{c|}{$\times$} &
  \multicolumn{1}{c|}{$\checkmark$} &
  \multicolumn{1}{c|}{$\times$} &
  $\checkmark$ \\ \hline
\multicolumn{3}{|l|}{Wang et al. \cite{wang2024make}} &
  \multicolumn{1}{c|}{---} &
  \multicolumn{1}{c|}{---} &
  \multicolumn{1}{c|}{---} &
  \multicolumn{1}{c|}{$\times$} &
  \multicolumn{1}{c|}{$\checkmark$} &
  \multicolumn{1}{c|}{$\times$} &
  \multicolumn{1}{c|}{$\checkmark$} &
  \multicolumn{1}{c|}{$\checkmark$} &
  \multicolumn{1}{c|}{$\checkmark$} &
  \multicolumn{1}{c|}{$\times$} &
  $\checkmark$ \\ \hline
\multicolumn{2}{|l|}{\multirow{3}{*}{Halim et al. \cite{blowprint}}} &
  Acoustic Input &
  \multicolumn{1}{c|}{99.35\%} &
  \multicolumn{1}{c|}{0.27\%} &
  \multicolumn{1}{c|}{19.2\%} &
  \multicolumn{1}{c|}{\multirow{3}{*}{$\times$}} &
  \multicolumn{1}{c|}{\multirow{3}{*}{$\times$}} &
  \multicolumn{1}{c|}{\multirow{3}{*}{$\times$}} &
  \multicolumn{1}{c|}{\multirow{3}{*}{$\checkmark$}} &
  \multicolumn{1}{c|}{\multirow{3}{*}{$\checkmark$}} &
  \multicolumn{1}{c|}{\multirow{3}{*}{$\checkmark$}} &
  \multicolumn{1}{c|}{\multirow{3}{*}{$\checkmark$}} &
  \multirow{3}{*}{$\checkmark$} \\ \cline{3-6}
\multicolumn{2}{|l|}{} &
  Facial Input &
  \multicolumn{1}{c|}{99.96\%} &
  \multicolumn{1}{c|}{0.04\%} &
  \multicolumn{1}{c|}{0\%} &
  \multicolumn{1}{c|}{} &
  \multicolumn{1}{c|}{} &
  \multicolumn{1}{c|}{} &
  \multicolumn{1}{c|}{} &
  \multicolumn{1}{c|}{} &
  \multicolumn{1}{c|}{} &
  \multicolumn{1}{c|}{} &
   \\ \cline{3-6}
\multicolumn{2}{|l|}{} &
  Score-Level Fusion &
  \multicolumn{1}{c|}{99.82\%} &
  \multicolumn{1}{c|}{0.18\%} &
  \multicolumn{1}{c|}{0\%} &
  \multicolumn{1}{c|}{} &
  \multicolumn{1}{c|}{} &
  \multicolumn{1}{c|}{} &
  \multicolumn{1}{c|}{} &
  \multicolumn{1}{c|}{} &
  \multicolumn{1}{c|}{} &
  \multicolumn{1}{c|}{} &
  \\ 
  \hlineB{3}
\multicolumn{2}{|l|}{\multirow{4}{*}{Ours}} &
  Blow-Acoustic &
  \multicolumn{1}{c|}{99.56\%} &
  \multicolumn{1}{c|}{0.30\%} &
  \multicolumn{1}{c|}{7.6\%} &
  \multicolumn{1}{c|}{\multirow{4}{*}{$\checkmark$}} &
  \multicolumn{1}{c|}{\multirow{4}{*}{$\checkmark$}} &
  \multicolumn{1}{c|}{\multirow{4}{*}{$\checkmark$}} &
  \multicolumn{1}{c|}{\multirow{4}{*}{$\checkmark$}} &
  \multicolumn{1}{c|}{\multirow{4}{*}{$\checkmark$}} &
  \multicolumn{1}{c|}{\multirow{4}{*}{$\checkmark$}} &
  \multicolumn{1}{c|}{\multirow{4}{*}{$\checkmark$}} &
  \multirow{4}{*}{$\checkmark$}\\ \cline{3-6}
\multicolumn{1}{|l}{} &
  \multicolumn{1}{c|}{} &
  Facial-Recognition &
  \multicolumn{1}{c|}{100\%} &
  \multicolumn{1}{c|}{0\%} &
  \multicolumn{1}{c|}{0\%} &
  \multicolumn{1}{c|}{} &
  \multicolumn{1}{c|}{} &
  \multicolumn{1}{c|}{} &
  \multicolumn{1}{c|}{} &
  \multicolumn{1}{c|}{} &
  \multicolumn{1}{c|}{} &
  \multicolumn{1}{c|}{} &
   \\ \cline{3-6}
\multicolumn{1}{|l}{} &
  \multicolumn{1}{c|}{} &
  Score-Level Fusion &
  \multicolumn{1}{c|}{99.95\%} &
  \multicolumn{1}{c|}{0.05\%} &
  \multicolumn{1}{c|}{0\%} &
  \multicolumn{1}{c|}{} &
  \multicolumn{1}{c|}{} &
  \multicolumn{1}{c|}{} &
  \multicolumn{1}{c|}{} &
  \multicolumn{1}{c|}{} &
  \multicolumn{1}{c|}{} &
  \multicolumn{1}{c|}{} &
   \\ \cline{3-6}
\multicolumn{1}{|l}{} &
  \multicolumn{1}{c|}{} &
  Feature-Level Fusion &
  \multicolumn{1}{c|}{100\%} &
  \multicolumn{1}{c|}{0\%} &
  \multicolumn{1}{c|}{0\%} &
  \multicolumn{1}{c|}{} &
  \multicolumn{1}{c|}{} &
  \multicolumn{1}{c|}{} &
  \multicolumn{1}{c|}{} &
  \multicolumn{1}{c|}{} &
  \multicolumn{1}{c|}{} &
  \multicolumn{1}{c|}{} &
   \\  \hline
  
\end{tabular}%
}
\smallskip\\
\begin{tabularx}{\textwidth}{X}
{\footnotesize
\textbf{Description of Notations}: R1: Accuracy, R2: Liveness Detection, R3: Revocability, R4: Resilience against known attacks, R5: Non-Invasiveness, R6: Usability, R7: MFB Support, R8: Low Resource Requirements. KR: Key Revocation, TR: Template Revocation, $\checkmark$: Achieved, $\times$: Not Achieved, ---: Not Reported}
\end{tabularx}
\end{table*}

\section{Conclusion}




This paper introduces BlowLive, a robust and highly accurate 
MFB framework that integrates blow-acoustic and facial biometrics as complementary behavioral and physiological modalities. The framework supports an FE-based authentication, which are designed to authenticate users through FE-derived cryptographic keys.
 By combining advanced spectral feature extraction with multimodal fusion, BlowLive achieves high authentication accuracy, exceeding $99.5\%$ in the behavioral modality and reaching $100\%$ in multi-factor settings. This design also supports template and key revocability in the event of compromise.

To counter spoofing attacks, we further propose 
a novel Doppler-shift-based liveness detection mechanism that systematically analyzes airflow turbulence generated during phone-blowing interactions. This approach effectively verifies the liveness of biometric inputs with 99.42\% accuracy, and helps defend against a wide range of playback and deepfake attacks. 

Overall, the proposed framework combines strong security guarantees, robust liveness detection, privacy-preserving authentication, and high usability within a practical multi-factor biometric framework.


\section*{Acknowledgement}
This research is supported by the National Research Foundation, Singapore and Infocomm Media Development Authority under its Trust Tech Funding Initiative (DTC-T2FI-CFP-0002). Any opinions, findings and conclusions or recommendations expressed in this material are those of the author(s) and do not reflect the views of National Research Foundation, Singapore and Infocomm Media Development Authority.

\bibliographystyle{splncs04}
\bibliography{10_references}


\end{document}